%% file: main.tex
\documentclass[a4paper,11pt]{article}

\input{macro}

\title{3-Local Hamiltonian Problem and Constant Relative Error  Quantum Partition Function Approximation:\\ $O(2^{\frac{n}{2}})$ Algorithm Is Nearly Optimal under QSETH}

\ifdefined\authorNote
\newcommand{\yc}[1]{{\color{magenta} [{YC:} #1]}}
\newcommand{\nai}[1]{{\color{brown} [{Nai:} #1]}}

\else
\newcommand{\yc}[1]{}
\newcommand{\nai}[1]{}
\fi

\newcommand{\email}[1]{\href{mailto:#1}{#1}}

\author{
    Nai-Hui Chia\thanks{Ken Kennedy Institute and Smalley-Curl Institute, Rice University, USA \email{nc67@rice.edu}.}  \and
    Yu-Ching Shen \thanks{Rice University, USA \email{ycshen@rice.edu}.}
}

\date{}

\begin{document}
\maketitle

\begin{abstract}
    We investigate the computational complexity of the Local Hamiltonian (LH) problem and the approximation of the Quantum Partition Function (QPF), two central problems in quantum many-body physics and quantum complexity theory. Both problems are known to be QMA-hard, and under the widely believed assumption that $\mathsf{BQP} \neq \mathsf{QMA}$, no efficient quantum algorithm exits. The best known quantum algorithm for LH runs in $O\bigl(2^{\frac{n}{2}(1 - o(1))}\bigr)$ time, while for QPF, the state-of-the-art algorithm achieves relative error $\delta$ in $O^\ast\bigl(\frac{1}{\delta}\sqrt{\frac{2^n}{Z}}\bigr)$ time, where $Z$ denotes the value of the partition function. A nature open question is whether more efficient algorithms exist for both problems. 

In this work, we establish tight conditional lower bounds showing that these algorithms are nearly optimal. Under the plausible Quantum Strong Exponential Time Hypothesis (QSETH), we prove that no quantum algorithm can solve either LH or approximate QPF significantly faster than $O(2^{n/2})$, even for 3-local Hamiltonians. In particular, we show:  
1) 3-local LH cannot be solved in time $O(2^{\frac{n}{2}(1-\varepsilon)})$ for any $\varepsilon > 0$ under QSETH;  
2) 3-local QPF cannot be approximated up to any constant relative error in $O(2^{\frac{n}{2}(1-\varepsilon)})$ time for any $\varepsilon > 0$ under QSETH; and  
3) we present a quantum algorithm that approximates QPF up to relative error $1/2 + 1/\poly(n)$ in $O^\ast(2^{n/2})$ time, matching our conditional lower bound.

Notably, our results provide the first fine-grained lower bounds for both LH and QPF with \emph{fixed locality}, namely for 3-local LH and 3-local QPF. 
This stands in sharp contrast to QSETH and the trivial fine-grained lower bounds for LH, where the locality of the SAT instance and the Hamiltonian depends on the parameter $\varepsilon$ in the $O(2^{\frac{n}{2}(1-\varepsilon)})$ running time. 
Our results align with the structure of most physical systems, where the number of particles involved in each interaction is fixed.

In summary, our lower and upper bounds suggest that there is little room for improving existing algorithms for LH and QPF with relative error greater than $1/2$, even when imposing locality constraints, assuming QSETH. 
This delineates a precise algorithmic barrier for these fundamental problems in quantum computing.
\end{abstract}

\tableofcontents
\input{1_intro}
\input{2_prelim}
\input{3_problem}
\input{4_0_lower_bound}
\input{5_algorithm}
\bibliographystyle{alphaurl}
\bibliography{References.bib}
\appendix
\input{A_SAT2H}

\end{document}

%% file: macro.tex
\usepackage[utf8]{inputenc}
\usepackage{amsmath,amssymb}
\usepackage{framed,caption,color,url}
\usepackage{standalone}
\usepackage{times}
\usepackage{mdframed}
\usepackage{algorithm2e,algorithmicx,algpseudocode,tikz,wrapfig}
\usetikzlibrary{decorations.pathreplacing,calligraphy,arrows, arrows.meta}
\usepackage{graphicx}               
\usepackage{cite} 
\usepackage{nicefrac} 
\usepackage{paralist} 
\usepackage{boxedminipage} 
\usepackage[pdfstartview=FitH,colorlinks,linkcolor=blue,filecolor=blue,citecolor=blue,urlcolor=blue,pagebackref=true]{hyperref}
\usepackage{upgreek}
\usepackage{centernot}
\usepackage{cleveref} 
\usepackage{subcaption}

\usepackage{bbm} 

\usepackage{fullpage}
\usepackage{amsthm}

\newcommand{\spn}{\mathrm{Span}}

\newcommand{\eg}{e.g.,\ }

\newcommand{\ceil}[1]{\lceil #1 \rceil}

\newcommand{\bra}[1]{\langle#1\rvert}
\newcommand{\ket}[1]{\lvert#1\rangle}

\newcommand{\inner}[2]{\langle#1\vert#2\rangle}

\usepackage{bm}

\newcommand{\poly}{\operatorname{poly}}

\newcommand{\negl}{\operatorname{negl}}

\newtheorem{theorem}{Theorem}[section]
\theoremstyle{plain}

\newtheorem{lemma}[theorem]{Lemma}
\newtheorem{corollary}[theorem]{Corollary}
\newtheorem{conjecture}[theorem]{Conjecture}

\theoremstyle{definition}

\newtheorem{definition}[theorem]{Definition}

\theoremstyle{definition}
\newtheorem{remark}[theorem]{Remark}

\newcommand{\tr}[1]{\mathrm{Tr}(#1)}
\newcommand{\braket}[2]{\inner{#1}{#2}}
\newcommand{\ketbra}[2]{\lvert#1\rangle\langle#2\rvert}
\newcommand{\proj}[1]{\ketbra{#1}{#1}}
\newcommand{\bracket}[3]{\left\langle#1\middle\rvert#2\middle\lvert#3\right\rangle}
\newcommand{\expec}[2]{\bracket{#1}{#2}{#1}}

\input{macrosymbols}

\newcommand{\tZ}{\widetilde{Z}}

\newcommand{\indicator}{\mathbbm{1}}

\usepackage{enumitem}

%% file: macrosymbols.tex
\newcommand{\rgst}[1]{\mathsf{#1}}
\newcommand{\reg}[1]{_\rgst{#1}}
\newcommand{\hs}[1]{\mathcal{#1}}
\newcommand{\notgate}{\textsc{not}}
\newcommand{\cnot}{\textsc{cnot}}

\newcommand{\addone}{\textsc{addone}}
\newcommand{\compare}{\textsc{compare}}

\newcommand{\zout}{\tZ} 

\newcommand{\clock}[1]{\gamma_{#1}}

\newcommand{\nclock}{n_{cl}}

%% file: 1_intro.tex
\section{Introduction}
Calculating the ground-state energy and the partition function is a fundamental task that appears in many fields, including quantum physics, quantum chemistry, and materials science~\cite{KLC24}. 
For instance, when developing a new compound, computing these quantities is essential for understanding the material's physical and chemical properties.
In these applications, systems are typically modeled by \emph{local Hamiltonians}.
Here, the term ``local'' means that the physical interaction involves only a small number of particles, and the overall Hamiltonian is the sum of all local interactions among the particles in the system\footnote{Although people are used to call it a local Hamiltonian, the physical interaction is not restricted in geometry. The interaction is allowed to be long-range.}.
For instance, the well-known Ising model \cite{SAC11} is a 2-local Hamiltonian.
\yc{Okay, I realize that I don't know how to describe a local Hamiltonian without using math formulas.}
Therefore, developing efficient algorithms for computing the ground-state energy and quantum partition function of local Hamiltonians is highly desirable.

Unfortunately, under the plausible conjecture that BQP $\neq$ QMA, no polynomial-time quantum algorithm is known for computing these quantities. 
The \emph{Local Hamiltonian problem (LH)} is the decision version of computing the ground-state energy. 
More formally, we are given two thresholds $a$ and $b$ with a promise gap $b - a = 1/\poly(n)$, where $n$ is the number of qubits (see Definition~\ref{def:LHP}). 
The input Hamiltonian is promised to have ground-state energy either at most $a$ or at least $b$, and the goal is to decide which case holds. 
Clearly, if we could compute the ground-state energy of any Hamiltonian to within a sufficiently small inverse-polynomial additive error, we could solve the corresponding LH decision problem by simply comparing the computed energy with the thresholds. 
However, it is well known that LH is QMA-complete: it is in QMA because, given the ground state, one can efficiently estimate its energy by measuring the Hamiltonian; for QMA-hardness, every problem in QMA can be reduced to an instance of LH~\cite{KSV02}. 
Therefore, the existence of a polynomial-time quantum algorithm for estimating the ground-state energy to inverse-polynomial accuracy would imply BQP $=$ QMA.

Computing the quantum partition function is an even harder problem. 
The partition function is a function of the temperature and the Hamiltonian of the system. 
It is defined by
\[
    Z:=\tr{e^{-\beta H}},
\]
where $\beta$ is the inverse of the temperature and $H$ is the Hamiltonian.
The \emph{Quantum Partition Function (QPF)} problem asks us to compute the value of $Z$ for a given Hamiltonian at a specified temperature. 
Intuitively, QPF is harder than LH because evaluating the partition function requires information about the \emph{entire} spectrum of the Hamiltonian, whereas solving LH only involves the ground-state energy. 
In fact, one can show that LH reduces to QPF, and thus QPF is QMA-hard (see \Cref{sec:fine-grained_reduction}). 
Moreover, QPF remains hard even for approximation. Bravyi et~al.~\cite{BCGW22} show that approximating the quantum partition function up to a relative error is computationally equivalent to approximately counting the number of witnesses accepted by a QMA verifier. Also, the reduction from LH to QPF still holds even for any constant relative error.

Although a polynomial-time algorithm does not exist under the assumption BQP $\neq$ QMA, algorithms better than the naive approach do exist. 
The naive way to compute the ground-state energy and the partition function is via full eigenvalue decomposition, which requires $O(2^{3n})$ time for an $n$-qubit Hamiltonian. 
In contrast, several quantum algorithms achieve substantially better performance. For the ground-state energy problem, there are quantum algorithms running in $O^\ast(2^{n/2})$ time\footnote{The $O^\ast(\cdot)$ notation hides polynomial factors in $n$.} for an $n$-qubit local Hamiltonian; 
Kerzner et~al.~\cite{KGM+24} propose a quantum algorithm that approximates the ground-state energy up to additive error $\varepsilon$ in $O^\ast\bigl(2^{n/2}/\varepsilon\bigr)$ time, and 
Buhrman et~al.~\cite{BGL+25} further improve this upper bound: their algorithm runs in $O^\ast\left(2^{\frac{n}{2}\left(1 - \frac{\varepsilon}{\varepsilon + k}\right)}\right)$
time for a $k$-local Hamiltonian\footnote{A $k$-local Hamiltonian is a sum of terms, each acting on at most $k$ particles.}. 
When the desired additive error is $1/\poly(n)$, the running time of the algorithm in~\cite{BGL+25} asymptotically approaches $O^\ast(2^{n/2})$ as $n \to \infty$. For the quantum partition function problem, Bravyi et~al.~\cite{BCGW22} propose a quantum algorithm that approximates the partition function $Z$ of an $n$-qubit system up to relative error $\delta$ in $O^\ast\left(\frac{1}{\delta}\sqrt{\frac{2^n}{Z}}\right)$ time. 
When the temperature is not too low (i.e., $\beta$ is not too large), $Z$ is not exponentially small, and the running time is roughly $O^\ast(2^{n/2})$.

In summary, all of these algorithms run in roughly $O^\ast(2^{n/2})$ time. 
This leads to the following natural question:
\begin{center}
    \emph{Do there exist algorithms that are significantly faster than $O(2^{n/2})$ for computing the ground-state energy and approximating the quantum partition function?}
\end{center}

\subsection{Our result}
\label{sec:our_result}
In this work, we give a \emph{negative answer} to the above question by ruling out the possibility of algorithms significantly faster than $O(2^{n/2})$ under a well-known complexity assumption. 
More precisely, we prove that, under the Quantum Strong Exponential Time Hypothesis (QSETH), neither solving the LH problem nor approximating the quantum partition function up to a constant relative error can be done in time $O\bigl(2^{\frac{n}{2}(1 - \varepsilon)}\bigr)$ for any constant $\varepsilon > 0$.

In particular, our first result establishes a lower bound for the Local Hamiltonian problem under QSETH.
\begin{theorem}[Informal version of \Cref{thm:main_lh} and \Cref{thm:main:3LH}]\label{thm:main_lh_informal}
    Assuming QSETH, the local Hamiltonian problem for 5-local and 3-local Hamiltonians cannot be solved by any quantum algorithm in time $O\bigl(2^{\frac{n}{2}(1-\varepsilon)}\bigr)$ for any constant $\varepsilon > 0$.
\end{theorem}

\begin{remark}
In \Cref{thm:main_lh_informal}, the operator norm of each local term in the 5-local Hamiltonian is upper bounded by 1, while the operator norm of each local term in the 3-local Hamiltonian is bounded by $\poly(n)$. 
\end{remark}

Our second result demonstrates a lower bound for approximating the quantum partition function up to any constant relative error under QSETH.
\begin{theorem}[Informal version of \Cref{thm:main_qpf} and \Cref{thm:main:3QPF}]
    \label{thm:main_qpf_informal}
    Assuming QSETH, approximating the quantum partition function for 5-local and 3-local Hamiltonians to within any constant relative error cannot be done by any quantum algorithm in time $O\bigl(2^{\frac{n}{2}(1-\varepsilon)}\bigr)$ for any constant $\varepsilon > 0$. 
\end{theorem}
\begin{remark}
It is worth noting that our reduction to 5-local Hamiltonians requires an inverse temperature $\beta$ that is asymptotically smaller than that required for the reduction to 3-local Hamiltonians. This can be interpreted as a \emph{temperature–locality tradeoff} between the two results.
\end{remark}

We give the definition of QSETH in the following.
\begin{definition}[QSETH, informal version of Conjecture \ref{conj:QSETH}]
    QSETH is a conjecture in which for all $\varepsilon>0$, there exists $k(\varepsilon)$ such that $k(\varepsilon)$SAT problem cannot be solved in $O(2^{\frac{n}{2}(1-\varepsilon)})$.
\end{definition}

QSETH is the quantum counterpart of the Strong Exponential Time Hypothesis (SETH). 
SETH conjectures that the Boolean satisfiability (SAT) problem (see Definition~\ref{def:kSAT} for the formal definition) on $n$ variables cannot be solved by any classical algorithm in time $O\bigl(2^{n(1-\varepsilon)}\bigr)$ for any constant $\varepsilon > 0$~\cite{IPZ99}. That is, a brute-force search is nearly optimal for SAT in the classical setting. As the quantum analogue of SETH, QSETH conjectures that SAT cannot be solved by any quantum algorithm in time $O\bigl(2^{\frac{n}{2}(1-\varepsilon)}\bigr)$ for any constant $\varepsilon > 0$~\cite{ACL+20,BPS21}. 
In other words, Grover’s search algorithm~\cite{Gro96} is essentially optimal for solving SAT on a quantum computer.

QSETH is regarded as a plausible conjecture in quantum computing and serves as a useful tool for deriving conditional lower bounds for various fundamental problems. 
First, the SAT problem has been studied for decades, and despite this extensive effort, the $\Omega(2^{n})$ (resp. $\Omega(2^{n/2})$) lower bound for classical (resp. quantum) algorithms has not been broken; designing quantum algorithms that refute these conjectures would likely require fundamentally new techniques and could lead to major breakthroughs in quantum algorithm design. 
In addition, QSETH has been used to establish conditional quantum time lower bounds for a wide range of problems, including the orthogonal vectors problem~\cite{ACL+20}, the closest pair problem~\cite{ACL+20,BPS21}, the edit distance problem~\cite{BPS21}, several lattice problems~\cite{CCK+23,HKW24}, and others~\cite{CCK+23}. 
These results highlight the value of QSETH as a powerful framework for studying the quantum hardness of computational problems.

Notably, Theorem~\ref{thm:main_lh_informal} and Theorem~\ref{thm:main_qpf_informal} are strong in the sense that relaxing locality and relative error do not help to make the problems easier. In particular, our lower bounds have the following properties that are distinct from standard QSETH.  
\begin{itemize}
    \item \textbf{Locality is independent of $\varepsilon$:} 
We emphasize that in our lower bounds for LH and QPF, the locality of the Hamiltonian is \emph{independent} of the parameter $\varepsilon$: for any $\varepsilon > 0$, there exists a family of 5-local and 3-local Hamiltonians that are hard to solve. 
In contrast, QSETH states that for all $\varepsilon > 0$, there exists a family of $k$-CNF Boolean formulas that are hard to solve, where the parameter $k$ depends on $\varepsilon$. 
(See Definitions~\ref{def:klocalH} and~\ref{def:kCNF} for the formal definitions of Hamiltonian locality and $k$-CNF formulas, respectively.) 
Having the locality remain independent of $\varepsilon$ strengthens our lower bound results, as it aligns with the physically realistic setting in which the number of particles participating in each interaction is fixed.
\item \textbf{Theorem~\ref{thm:main_qpf_informal} holds for any constant relative error:} 
The algorithm for QPF is required to output an approximation $\zout$ such that 
\[
    (1 - \delta)\zout \le Z \le (1 + \delta)\zout,
\]
where $\delta \in (0,1)$ is the relative error parameter. 
Intuitively, the problem should become easier as $\delta$ increases. 
However, our result shows that even if $\delta$ is allowed to be any constant, QPF still cannot be solved in time $O\bigl(2^{\frac{n}{2}(1 - \varepsilon)}\bigr)$ for any constant $\varepsilon > 0$.

\end{itemize}

Our lower bound for the Local Hamiltonian problem is $\Omega\bigl(2^{\frac{n}{2}-o(1)}\bigr)$, which matches the performance of the best known quantum algorithms running in $O^\ast\bigl(2^{\frac{n}{2}-o(1)}\bigr)$ time~\cite{BGL+25}. 
This yields the following corollary.

\begin{corollary}\label{cor:lh_ub_informal}
    Any $O^\ast\bigl(2^{\frac{n}{2}(1 - o(1))}\bigr)$-time quantum algorithm for computing the ground-state energy of local Hamiltonians, such as the one in~\cite{BGL+25}, is optimal for the Local Hamiltonian problem under QSETH.
\end{corollary}

\begin{remark}
At first glance, our 
$\Omega\bigl(2^{\frac{n}{2}(1-\varepsilon)}\bigr)$ lower bound for any $\varepsilon > 0$ appears to contradict the $O^\ast\bigl(2^{\frac{n}{2}(1-\varepsilon')}\bigr)$ upper bound for estimating the ground-state energy up to an additive error $\varepsilon'$ established in~\cite{BGL+25}. 
The key point is that solving the LH problem requires estimating the ground-state energy to within an additive error of $1/\poly(n)$. 
Consequently, the running time of the algorithm in~\cite{BGL+25} is 
$O^\ast\left(2^{\frac{n}{2}\left(1 - \frac{1}{\poly(n)}\right)}\right)$.
For any fixed $\varepsilon > 0$, we have $2^{\frac{n}{2}\left(1 - \frac{1}{\poly(n)}\right)} > 2^{\frac{n}{2}(1 - \varepsilon)}$
when $n$ is larger than some constant $n_0$. 
Therefore, the lower and upper bounds are consistent.
\end{remark}

Finally, we propose a quantum algorithm for approximating QPF whose running time matches the $\Omega(2^{\frac{n}{2}(1-o(1))})$ lower bound.
\begin{theorem}[Informal version of \Cref{thm:main_qpf_alg}]
    \label{thm:main_qpf_ub_informal}
    There exists a quantum algorithm whose running time is $O^{\ast}(2^\frac{n}{2})$ and the algorithm approximates the quantum partition function for local Hamiltonians up to a relative error $\frac{1}{2}+\frac{1}{\poly(n)}$,
    where $n$ is the number of qubits of the Hamiltonian.
\end{theorem}
We now compare our algorithm with the one proposed in~\cite{BCGW22}. 
The running time of their algorithm is $O^\ast\left(\frac{1}{\delta}\sqrt{\frac{2^n}{Z}}\right)$,
which depends on the value of the partition function $Z$. 
When the inverse temperature $\beta = O(n^c)$ for some large constant $c$ (i.e., at low temperatures), the running time may exceed $O(2^{n/2})$. 
In contrast, the running time of our algorithm does not depend on $Z$: it runs in $O^\ast(2^{n/2})$ for all $\beta = \poly(n)$.

There are also differences in the accuracy and space requirements. 
The algorithm in~\cite{BCGW22} works for relative errors $\delta \in 1/\poly(n)$, whereas our algorithm currently applies only to constant relative errors $\delta > 1/2$. 
In terms of space, their algorithm is highly space-efficient, requiring only $O\bigl(\log n + \log(1/\varepsilon)\bigr)$ ancilla qubits. 
By contrast, our algorithm uses phase estimation combined with the median-of-means technique, which requires a $\poly(n)$ number of ancilla qubits.

\subsection{Technical overview}
\subsubsection*{Lower bound for 5-local LH and QPF problems}
Our goal is to reduce $k$-SAT to the 5-local Hamiltonian problem. 
Let $\Phi$ be an instance of $k$-SAT defined on $n$ variables. 
Given $\Phi$, the reduction constructs a 5-local Hamiltonian $H$ acting on $n'$ qubits such that the ground state and ground-state energy of $H$ encode the information of $\Phi$. 
If the reduction can be performed in time $O\bigl(2^{\frac{n}{2}(1-\varepsilon)}\bigr)$, then QSETH would be violated. 
To achieve a lower bound of $\Omega\bigl(2^{\frac{n'}{2}(1-\varepsilon')}\bigr)$ for the Local Hamiltonian problem, matching the lower bound in QSETH, we need the reduction to be \emph{size-preserving}. 
Moreover, to obtain a fixed locality for $H$, the locality of $H$ must be independent of $k$, since $k$ depends on the parameter $\varepsilon$ in QSETH. In summary, the reduction must satisfy the following two conditions:  
\begin{enumerate}[label=(\arabic*)]
    \item \emph{Size-preserving}: $n' = n + o(n)$, and
    \item \emph{Locality-independent}: the locality of the resulting Hamiltonian does not depend on $k$.
\end{enumerate}

To obtain a size-preserving reduction, a straightforward approach is to directly construct $H$ from the $k$-SAT instance $\Phi$. 
Each variable in $\Phi$ corresponds to a qubit in $H$, and each clause in $\Phi$ is translated into a local term of $H$ (see \Cref{sec:appexdixA} for details). 
In this construction, the number of qubits in $H$ is $n$, and since each clause involves at most $k$ variables, each local term of $H$ acts on at most $k$ qubits. 
Therefore, $H$ is a $k$-local Hamiltonian acting on $n$ qubits, satisfying condition (1) (size preservation). 
However, the locality of $H$ matches $k$, which violates condition (2) (locality independence).

To satisfy condition (2), one can instead apply a circuit-to-Hamiltonian reduction, such as Kitaev et al.’s 5-local Hamiltonian construction~\cite{KSV02}. 
In this framework, any quantum circuit $U$ can be encoded into a 5-local Hamiltonian $H$. 
The key idea is to map a polynomial-time verification circuit into a 5-local Hamiltonian instance: if there exists a witness that causes the verification circuit to accept with high probability, then the corresponding Hamiltonian has low ground-state energy; otherwise, its ground-state energy is large.

However, Kitaev's reduction does not satisfy (1) since it increases the size of the Hamiltonian by the number of gates in the verification circuit. To be more specific, the reduction requires to record the computation process in the ground state of $H$ using a so-called \emph{clock state}, which indicates the progress of the computation. Notably, if $U$ consists of $g$ gates, the clock state register requires to have $g$ qubits. Together with the $n$ input qubits and $n_a$ ancilla qubits, the constructed Hamiltonian acts on $n' = n + n_a + g$ qubits. Since verifying an assignment on all $m$ clauses requires at least $m$ steps, the number of gates $g$ is $O(m)$. When $m$ is superlinear in $n$, the reduction fails. Even if we assume $m$ is linear in $n$, the number of qubits in the clock state is $cn$ for some constant $c>1$, so it does not satisfy condition (1).

To ensure that our reduction works, we need a construction of the clock state associated with the induced Hamiltonian that satisfy the following three conditions: 
\begin{enumerate}[label=(\roman*)]
    \item The clock state uses $o(n)$ of qubits and can encode $\poly(n)$ of computation steps,
    \item the operation on the clock state needs to be constant-local, and the locality is independent of the circuit, and
    \item the computation process of $U$ is encoded in the ground state of the induced Hamiltonian $H$.  
\end{enumerate}

In Kitaev’s reduction, conditions (ii) and (iii) are satisfied, but condition (i) is not, as discussed above. 
In~\cite{CCH+23}, a clock construction is proposed that satisfies conditions (i) and (ii). 
However, unlike Kitaev’s approach, the circuit-to-Hamiltonian reduction in~\cite{CCH+23} encodes the computation in the \emph{time evolution} of $H$, rather than in its ground state. As a result, there is no guarantee regarding the structure or properties of the ground state in their construction.

We build on the clock-state construction from~\cite{CCH+23} and the Hamiltonian construction from~\cite{KSV02}. 
By introducing a carefully designed penalty term into $H$, we enforce that the computation history is stored in the ground state. 
This yields a novel circuit-to-Hamiltonian reduction that simultaneously preserves the size of the Hamiltonian, maintains constant locality, and ensures that the circuit $U$ is encoded in the ground state of the resulting Hamiltonian $H$. 
Our construction satisfies all three requirements, overcoming the limitations that neither of the existing reductions can address individually.

Our clock-state construction guarantees that, as long as the number of gates in the verification circuit is polynomial, the clock-state register requires only $o(n)$ qubits. 
Consequently, if the verification circuit uses $o(n)$ ancilla qubits, the Hamiltonian $H$ produced by our circuit-to-Hamiltonian reduction acts on $(n + o(n))$ qubits and has constant locality. This satisfies both conditions (1) and (2).

The remaining task is to construct a verification circuit that uses $o(n)$ ancilla qubits for any $k$-SAT instance. 
To achieve this, we use a counter to record the number of clauses satisfied by a given assignment. 
The $k$-SAT formula is satisfied if and only if the value stored in the counter equals $m$. 
Since recording up to $m$ requires only $\log m$ qubits, the number of ancilla qubits needed is $n_a = O(\log m) = o(n)$, as required.

The lower bound for QPF follows directly from the lower bound for LH. 
When $\beta$ is a sufficiently large polynomial, the partition function is dominated by low-energy states. 
If the ground-state energy is at most $a$, then by ignoring all other states except the ground state, we obtain $Z \ge e^{-\beta a}$. 
Conversely, if the ground-state energy is at least $b$, then by treating all eigenstates as having energy $b$, we get $Z \le 2^n e^{-\beta b}$. 
Therefore, by approximating $Z$, we can distinguish between the two cases and decide the ground-state energy. 
To ensure this distinction, it suffices that $(1 - \delta) e^{-\beta a} > (1 + \delta) 2^n e^{-\beta b}$. 
In our reduction, we have observed that for any $\delta > 0$, this inequality holds for all sufficiently large $n$.

\subsubsection*{Lower bound for 3-local LH and QPF problems}
In our construction of clock state, we need to specify two parameter $\nclock,d\in\mathbb{N}$.
The integer $\nclock$ is the number of qubits in the clock register.
We require $\nclock^d\ge g$ to encode the computation step $1,2,\dots,g$.
Based on the QSETH assumption, $d=2$ suffices.

The locality 5 in the previous section comes from the fact that in our circuit to Hamiltonian reduction, we need a local term operating on $d+1$ qubits in the clock register and 1 or 2 qubits in the circuit register.
Hence, the induced Hamiltonian $H$ is $d+3$-local.
When $d=2$, we get a 5-local Hamiltonian.

We can further reduce the locality to 3-local by using the technique in \cite{KKR04}.
The key ideas in \cite{KKR04} are the following two.
\begin{enumerate}
    \item \textbf{1-local operator on the circuit register suffices.}  
    The reason is that we make the two-qubit gates in $U$ be control-$Z$ gates, and each control-$Z$ gate is preceded by two $Z$ gates, and followed by two $Z$ gates as well.
    Each $Z$ gate preceded by the control-$Z$ acts on one of qubits of the control-$Z$; and each $Z$ gate followed by the control-$Z$ acts on one of qubits of the control-$Z$ as well.
    By this circuit structure, we can use 1-local operators to encode the computation of the circuit. 
    This structure is without loss of generality because the control-$Z$ gate and single qubit gates form a universal gate set,
    and a control-$Z$ gate ``sandwiched'' by four $Z$ gates described above is identical to a control-$Z$ gate.
    \item \textbf{For the operators acting on the clock register, the locality can be reduced compared to~\cite{KSV02}.} 
This improvement comes from the projection lemma (Lemma~\ref{lem:projection_lemma}), which applies a sufficient ``penalty'' on any state that is not a valid clock state. 
As a result, when constructing the Hamiltonian for propagation, we only need to account for legal clock states. 
More specifically, the propagation Hamiltonian between times $t$ and $t'$ only needs to act on the qubits that differ between the $t$th and $t'$th clock states. 
This observation allows us to reduce the locality of the propagation Hamiltonian on the clock register.
\end{enumerate}

Ideally, we would hope that the two techniques in~\cite{KKR04} could reduce the locality of our Hamiltonian to $2$. 
However, in our clock configuration, transitioning from time $t$ to $t'$ requires changing at least two qubits in the clock register. 
Moreover, since $t'$ can be $t+2$ in the construction of~\cite{KKR04}, this leads to Hamiltonians that act on at least four qubits in our clock configuration. 
As a result, the locality of the Hamiltonian cannot be reduced to even $3$, as one might initially expect. 
To address this, we carefully design a scheduling for our clock-state configurations so that each transition, either one step ($t \to t+1$) or two steps ($t \to t+2$), changes exactly two qubits in the clock register. 
This construction yields a 3-local Hamiltonian, as stated in Theorem~\ref{thm:main_lh_informal}.

\subsubsection*{Upper bound for constant-local QPF problem}
We use the idea from the proof of the equivalence between approximating the number of witnesses of a verifier and approximating the quantum partition function in~\cite{BCGW22}. 
Briefly speaking, we divide the energy spectrum into polynomially many intervals, select a representative energy value $E_j$ from each interval $j$, count the number of eigenstates in each interval $j$, denoted by $M_j$, and then approximate the partition function $Z$ by 
$\sum_j M_j e^{-\beta E_j}$.

We now explain how to count $M_j$. 
First, we construct an energy estimation circuit $U_{EE}$ that outputs the corresponding eigenvalue $E_p$ for each eigenstate $\ket{\psi_p}$ of $H$. 
In other words, $U_{EE}\ket{\psi_p} = \ket{E_p}\ket{\psi_p}$. 
The circuit $U_{EE}$ can be implemented using Hamiltonian simulation together with phase estimation. Then, we apply a circuit $U_{dec}$ that decides whether $E_p$ is in the interval $j$.
Let $U_j:=U_{dec}U_{EE}$. Together, we have $U_j\ket{\psi_p}=\ket{1}\ket{\phi_p}$ if $E_p$ in the interval $j$ and $U_j\ket{\psi_p}=\ket{0}\ket{\phi_p}$ otherwise, where $\ket{\phi_p}$ is some quantum state corresponding to $\ket{\psi_p}$. 

Now, suppose we can prepare the uniform superposition of all eigenstates. Then, by applying $U_j$ on the state, we obtain
\begin{equation}
    \label{eq:tech_count}
    U_j\sum_{p}\frac{1}{\sqrt{N}}\ket{\psi_p} = \sqrt{\frac{M_j}{N}}\ket{1}\ket{\xi_1} + \sqrt{\frac{N-M_j}{N}}\ket{0}\ket{\xi_0},
\end{equation}
where $N=2^n$ is the number of the eigenstates and $\ket{\xi_1}$, $\ket{\xi_0}$ are two quantum states orthogonal to each other. Finally, by using the well-known amplitude estimation algorithm on \Cref{eq:tech_count}, we can obtain $M_j$. 

However, directly preparing the uniform superposition of all eigenstates is generally computationally hard  
since the eigenstates are unknown. To overcome this difficulty, we use the idea introduced in~\cite{KGM+24}. 
The key observation is that the uniform superposition over a complete basis, tensored with its complex conjugate, is equivalent to an EPR state. 
Therefore, by applying $U_j$ to the EPR state, we obtain
\begin{equation}
    \label{eq:tech_count_epr}
    U_j \sum_{p} \frac{1}{\sqrt{N}}\ket{\psi_p}\ket{\psi_p^{\ast}}
    = \sqrt{\frac{M_j}{N}}\ket{1}\ket{\xi'_1} 
    + \sqrt{\frac{N - M_j}{N}}\ket{0}\ket{\xi'_0},
\end{equation}
where $\ket{\xi'_1}$ and $\ket{\xi'_0}$ are orthogonal quantum states. 
By applying the amplitude estimation algorithm to the state in~\Cref{eq:tech_count_epr}, we can approximate $M_j$.

\subsection{Open questions}
We summarize our results and list some open questions in the following.

\paragraph{Local Hamiltonian problem.}
We establish an $\Omega(2^{\frac{n}{2}(1-o(1))})$ lower bound for the Local Hamiltonian (LH) problem on 3-local Hamiltonians, matching the complexity of the best known algorithms. This result suggests that incorporating locality into algorithm design is unlikely to yield significant speedups, except possibly for the 2-local case. A natural open question is therefore: \emph{Can we reduce the locality to 2-local?} A main barrier in our construction lies in the need to use a 2-local operator to update the clock state. When combined with the operators describing the computational circuit, the overall locality exceeds two. On the other hand, existing algorithms for the LH problem with inverse-polynomial promise gap typically do not exploit locality. Thus, it remains possible that there exist algorithms substantially faster than $2^{n/2}$ for 2-local Hamiltonians.

Also, our lower bound holds when there is no restriction on the geometry configuration of the Hamiltonian. However, many physical systems in nature exhibit constrained geometries. It is known that both two-dimensional and one-dimensional qudit systems are QMA-complete~\cite{OT08,AGIK09,HNN13}. This leads to a natural question: \emph{Assuming QSETH, does the $\Omega(2^{\frac{n}{2}(1-o(1))})$ lower bound still hold for geometrically local Hamiltonians?} 

Furthermore, one can ask whether the lower bound continues to hold when each local term is restricted to a specific form. For example, in the quantum Max-Cut problem~\cite{GP19}, each local term of the Hamiltonian takes the form $I-X\otimes X - Y\otimes Y -Z\otimes Z$, and the problem is still QMA-complete. 

Finally, our threshold gap $b-a$ in the 5LH problem is $O(1/n^6)$, and the number of terms in the 5-local Hamiltonian is $m=O(n^3)$ (see \Cref{sec:5local}).
This gives a relative gap $(b-a)/m=O(1/n^9)$.
A natural question arises: when the relative gap increases, for example, $(b - a)/m = O(1)$, does the $\Omega(2^{\frac{n}{2}(1-o(1))})$ lower bound still hold under QSETH?

\paragraph{Quantum partition function problem} In this work, we show that LH can be reduced to QPF, and the reduction is fine-grained even for an arbitrary constant relative error; therefore, we can obtain the $\Omega(2^{n/2})$ lower bound assuming QSETH. Notably, the lower bound holds for any constant relative error and locality at least $3$. Along this line, we can ask whether the lower bound continues to hold when considering 2-local Hamiltonian, geometrical restrictions, and specified local terms, or we can design faster algorithms under these constraints.  

Furthermore, another interesting question is whether the reverse reduction holds, namely, whether approximating QPF can be reduced to the LH problem. At first glance, this seems unlikely, as QPF can be interpreted as estimating the dimension of the ground space in the low-temperature regime. However, since we are considering exponential lower bounds, such a reduction could, in principle, be allowed to run in super-polynomial time.

Our lower bound for the 5QPF problem holds for $\beta = O(n^7)$ and $\|H\| = O(n^3)$, while the operator norm of the corresponding 3-local Hamiltonian is even larger. A natural question to ask is what happens as the temperature increases. In other words, when $\beta\|H\|$ becomes small, does the lower bound still hold in this regime? Or, at what point do we observe a \emph{phase transition}?

The state-of-the-art algorithm for QPF achieves accuracy for any inverse-polynomial relative error, but it does not guarantee an $O^{\ast}(2^{\frac{n}{2}})$ running time for arbitrary $k$-local Hamiltonians~\cite{BCGW22}.
In contrast, our $k$QPF algorithm promises an $O^{\ast}(2^{\frac{n}{2}})$ running time but only for constant relative error $\delta > 1/2$. This raises the following question: \emph{Is there an $O^{\ast}(2^{\frac{n}{2}})$ algorithm for any constant relative error $\delta = O(1)$?}

\subsection{Acknowledgment}
We thank Chunhao Wang and Christopher Ye for valuable discussions.
Nai-Hui Chia is supported by NSF Award FET-2243659, NSF Career Award FET-2339116, Google Scholar Award, and DOE Quantum Testbed Finder Award DE-SC0024301. Yu-Ching Shen is supported by NSF Award FET-2339116 and NSF Award FET-2243659.

%% file: 2_prelim.tex
\section{Preliminaries}
\subsection*{Notation}
For $n\in\mathbb{N}$, we use $[n]$ to denote the set collecting all the positive integers smaller than or equal to $n$, that is, $\{1,2,\dots,n\}$. 

For a bit string $x\in\{0,1\}^{*}$, we use $int(x)$ to denote the corresponding integer whose binary representation is $x$.
For a nonnegative integer $n$, we use $bin(n)$ to denote the binary representation of $n$.
The length of $bin(n)$ depends on the context. \yc{Well, this sentence sounds weird.}
For example, $int(01010111)=87$ and $bin(87)=010101111$.

For an $n$-bit string $x\in\{0,1\}^{n}$ and $i\in[n]$, we use $x[i]$ to denote the $i$th bit of $x$.
If $x$ is a binary representation of an integer, then $x[1]$ is the most significant bit and $x[n]$ is the least significant bit. 
Let $S\in[n]$, we use $x_S$ to denote a new string that concatenates $x$'s bits whose indices are in $S$.
For example, if $x=01010111$ and $S=\{1,3,7,8\}$, then $x_S=0011$.
The Hamming weight of $x$ is denoted by $wt(x)$, that is, $wt(x)$ is the number of 1's in $x$. 

The identity operator is denoted by $I$.
The Kronecker delta $\delta_{i,j}$ is defined by $\delta_{i,j}=1$ if $i=j$ and $\delta_{jk}=0$ if $i\neq j$. 
The indicator function is denoted by $\indicator_S(i)$, which is defined by $\indicator_S(i)=1$ if $i\in S$ and $\indicator_S(i)=0$ if $i\notin S$.

The big $O$ star notation $O^*(\cdot)$ hides the polynomial factors in the standard big $O$ notation.
For example, $8\cdot 2^\frac{n}{2}\cdot n^7\in O^*(2^{\frac{n}{2}})$.
The negligible functions $\negl(\cdot)$ are functions that are smaller than any inverse polynomial: if $\mu(n)\in\negl(n)$ then for all $c\in\mathbb{N}$,  there exists $n_0\in\mathbb{N}$ such that for all $n\ge n_0$, it holds that $\mu(n) < \frac{1}{n^c}$.

\subsection*{Quantum computation}
\paragraph{Quantum register}
We use the sans-serif font to denote quantum registers, \eg $\rgst{A}$, $\rgst{in}$, $\rgst{out}$, $\rgst{anc}$, $\rgst{clock}$.
Throughout this paper, a register consists of qubits.
A qubit may belong to different registers at the same time.
For example, a qubit in the output register $\rgst{out}$ of a quantum circuit is also in the ancilla register $\rgst{anc}$.
By an abuse of notation, sometimes we treat the resisters as sets.
We say $\rgst{A}\subseteq\rgst{B}$ if any qubit in $\rgst{A}$ is also a qubit in $\rgst{B}$.
We use $\rgst{A}\cup\rgst{B}$ to denote the register that consists of the qubits that are in $\rgst{A}$ or in $\rgst{B}$, $\rgst{A}\cup\rgst{B}$ to denote the portion that belongs to $\rgst{A}$ and $\rgst{B}$ at the same time, and $\rgst{A}\setminus\rgst{B}$ to denote the portion that belongs to $\rgst{A}$ but not $\rgst{B}$.
When we specify a system (that can be a quantum circuit or a physical qubit system), we use $\overline{\rgst{A}}$ to denote the collection of qubits in the system that are not in $\rgst{A}$.
We use $|\rgst{A}|$ to denote the number of qubits in $\rgst{A}$.
When we specify a register $\rgst{A}$, we use $\rgst{A[i]}$ to denote the $i$th qubit in $\rgst{A}$ for $i\in[|\rgst{A}|]$.

We use $\ket{\psi}\reg{a}$ to denote the register $\rgst{a}$ is in the state $\ket{\psi}$. 
For a Hermitian or a unitary operator $X$, we use $X\reg{A}$ to denote $X$ acting on the register $\rgst{A}$.
When the registers are specified, the order of the tensor product of the operators is not sensitive. For example, $X\reg{A}\otimes Y\reg{B} = Y\reg{B}\otimes X\reg{A}$.
If $\rgst{A}$ is a portion of the whole system, we say $X$ non-trivially acts on $\rgst{A}$ if the operator acting on the entire system is $X\reg{A}\otimes I\reg{\overline{A}}$.
We use $X^{\otimes k}$ to denote the $k$th tensor power of $X$ for $k\in\mathbb{N}$. Namely, $X^{\otimes k}:=\underbrace{X\otimes X\otimes\cdots\otimes X}_{k}$.
Hence, $X^{\otimes|\rgst{A}|}\reg{A}$ denotes that every qubit in the register $\rgst{A}$ is applied by $X$.

\paragraph{Quantum circuit} 
A quantum circuit is a unitary acting on the union of two disjoint register $\rgst{in}$ and $\rgst{anc}$, which are called input register and ancilla register respectively.
Also, there is an output register $\rgst{out}\subseteq\rgst{in}\cup\rgst{anc}$.
We require the ancilla register is $\ket{0^{n_a}}\reg{anc}$ initially, where $n_a=|\rgst{anc}|$.
That is, if the circuit $U$ takes a quantum state $\ket{\psi}$ as an input, then the final state of the circuit is $U\ket{\psi}\reg{in}\ket{0^{n_a}}\reg{anc}$.
We use the notation $x\gets U(\ket{\psi})$ to denote the event that we obtain the measurement outcome $x\in\{0,1\}^{|\rgst{out}|}$ when we measure on the output register $\rgst{out}$ at the end of the circuit $U$ that takes $\ket{\psi}$ as the input state.
We have $\Pr[x\gets U(\ket{\psi})]=\bra{\psi}\reg{in}\bra{0^{n_a}}\reg{anc}U^\dagger (\proj{x}\reg{out}\otimes I\reg{\overline{out}})U\ket{\psi}\reg{in}\ket{0^{n_a}}\reg{anc}$.

\paragraph{Universal gate set and elementary gates}
A quantum circuit is implemented by a sequence of quantum gates.
There is a finite set of one-qubit and two-qubit quantum gates called the universal gate set such that for any unitary $U$, we can use the gates in the universal gate sets to implement a quantum circuit arbitrarily close to $U$. 
We call a member in the universal gate set an elementary gate. 
We choose $\{\textsc{hadamard}, \pi/8\textsc{-gate}, \notgate, \cnot\}$ as the universal gate set \cite{NC10}.
The $\notgate$ gate acts on a one-qubit register.
The truth table of the $\notgate$ gate is defined by $\notgate\ket{x}=\ket{(1+x)\mod 2}$ where $x\in\{0,1\}$.

We also introduce a multi-control-NOT gate.
\begin{definition}[Multi-control-NOT gate and Toffoli gate]
    A multi-control gate, denoted by $C^k\notgate$, is a quantum gate that acts on $k+1$ qubits.
    Let $\rgst{C}$ be a $k$-qubit register and $\rgst{T}$ be a one-qubit register.
    The truth table of the $C^k\cnot$ gate is defined by 
    \[
    C^k\cnot\ket{x_1,x_2,\dots,x_k}\reg{C}\ket{x_{k+1}}\reg{T}=\ket{x_1,x_2,\dots,x_k}\reg{C}\ket{(x_1x_2\cdots x_k+x_{k+1})\mod 2}\reg{T},
    \]
    where $x_1,x_2,\dots,x_{k+1}\in\{0,1\}$.
    We call $\rgst{C}$ the control qubits and $\rgst{T}$ the target qubit.
    
    For $k=2$, we call $C^2\notgate$ a Toffoli gate.
\end{definition}

According to the definition of $C^k\notgate$, we have that  $\cnot$ is a special case of $C^k\notgate$ gate for $k=1$.

\begin{remark}[Decompose $C^k\notgate$ into Toffolis \cite{NC10}]
    A $C^k\notgate$ gate can be constructed by $2k-3$ Toffoli gates associated with $k-2$ ancilla qubits. The ancilla qubits are in all zero state initially and stay unchanged at the end of circuit.
\end{remark}

\begin{remark}[Decompose Toffoli into elementary gates\cite{NC10}]
    A Toffoli gate is identical to a circuit that acts on three qubits and is composed of 17 elementary gates.
\end{remark}

\subsection*{Hamiltonian} 
A Hamiltonian is an Hermitian operator acting on a quantum register.
Let $H$ be a Hamiltonian, we call the eigenvalues of $H$ the energies of the Hamiltonian.
We use $\lambda(H)$ to denote the smallest eigenvalue of $H$. 
We call $\lambda(H)$ the ground state energy of $H$, and we call the corresponding eigenstate(s) the ground state(s) of $H$. 
We use $\|H\|$ to denote the operator norm of $H$,
that is, the largest absolute value of the eigenvalue of $H$.

The $k$-local Hamiltonian is defined as follows.
\begin{definition}[Local Hamiltonian]\label{def:klocalH}
We say a Hamiltonian $H$ is $k$-local if $H$ can be written as $H = \sum_{i} H_i$ and for all $i$, the following holds.
\begin{itemize}
    \item $H_i$ is a Hamiltonian.
    \item $H_i$ non-trivially acts on at most $k$ qubits.
    \item $\|H_i\|\le \poly(n)$.
\end{itemize} 
\end{definition}
We call $k$ the locality of $H$. 

\subsection*{Boolean formula}
A Boolean variable can be assigned to the value 0 or 1.
We abbreviate the term Boolean variable as variable.
A Boolean formula consists of variables associated with parentheses and logic connectives $\neg$ (NOT), $\vee$ (OR), and $\wedge$ (AND).

Let $\Phi$ be a Boolean formula with $n$ variables $x_1,x_2,\dots,x_n$.
We use an an $n$-bit string $x\in\{0,1\}^n$ to denote an assignment to $x_1,x_2,\dots,x_n$. 
The variable $x_i$ for $i\in[n]$ is assigned to the $i$th bit of $x$.
We use $\Phi(x)$ to denote the value of $\Phi$ when  $x_1,x_2,\dots,x_n$ are assigned to $x$. 
We say $x$ satisfies $\Phi$ if $\Phi(x)=1$.

For all variables $x$, it holds that $\neg(\neg x)= x$.
Let $x_1, x_2,\dots,x_n$ be variables and for each $i\in[n]$, let $\ell_i$ be either $x_i$ or $\neg x_i$ it holds that $\neg(\ell_1\wedge \ell_2\wedge\cdots\wedge \ell_m) = \neg \ell_1 \vee \neg \ell_2 \vee\cdots\vee \neg \ell_m$.

We can use a quantum circuit to compute Boolean formulas.
For $x\in\{0,1\}$, it holds that
\begin{equation}\label{eq:q_comp_not}
    \notgate\ket{x} = \ket{\neg x}.
\end{equation}
For $x_1,x_2,\dots,x_k\in\{0,1\}$, it holds that
\begin{equation}\label{eq:q_comp_and}
    C^k\notgate\reg{C\cup T}\ket{x_1,x_2,\dots,x_k}\reg{C}\ket{0}\reg{T} = \ket{x_1,x_2,\dots,x_k}\reg{C}\ket{x_1\wedge x_2 \wedge \cdots\wedge x_k}\reg{T}.
\end{equation}

We define the conjunctive normal form formula as follows.
\begin{definition}[Conjunctive normal form (CNF) formula]\label{def:kCNF}
Let $\Phi$ be a Boolean formula that consists of $n$ variables $x_1,x_2,\dots, x_k$.
We say $\Phi$ is a $k$CNF formula if $\Phi$ is in the form of $m=\poly(n)$ smaller formulas connected by $\wedge$.
Each smaller formula contains at most $k$ variables, and the variables are connected by $\vee$.

To be more precise, $\Phi=\varphi_1\wedge \varphi_2 \wedge\cdots \wedge \varphi_m$ where $m=\poly(n)$, and for all $i\in [m]$, the following constrains hold.
\begin{itemize}
    \item $\varphi_{i} = (\ell_{i,1}\vee \ell_{i_2}\vee\cdots\vee\ell_{i,k_{i}})$, where $\ell_{i,p}$ can be $x_j$ or $\neg x_j$ for all $p\in[k_i]$ and $j\in[n]$.
    \item $k_i\le k$.
    \item For each $j\in[n]$, $x_j$ and $\neg x_j$ do not appear in $\varphi_i$ at the same time;
    $x_j$, $\neg x_j$ appears in $\varphi_i$ at most once.
\end{itemize}
We say $\varphi_i$ is a clause of $\Phi$ for all $i\in[m]$.
\end{definition}

%% file: 3_problem.tex
\section{Hamiltonian problems and fine-grained complexity}
\label{sec:problems}

\subsection{Local Hamiltonian and quantum partition problem}
\label{sec:lh_qpf}
In this section, we formally define the local Hamiltonian problem and the approximating quantum partition function problem.

\begin{definition}[$k$-local Hamiltonian ($k$LH) problem]
    \label{def:LHP}
    The local Hamiltonian problem is a decision problem that asks whether the ground state energy of a $K$-local Hamiltonian is greater or less than given energy thresholds.
    \begin{itemize}
        \item \textbf{Inputs:} a $k$-local Hamiltonian $H$ acting on $n$ qubits where $k=O(1)$, and two energy thresholds $a,b$ satisfying $b-a\ge 1/\poly(n)$.
        \item  \textbf{Outputs:}
        \begin{itemize}
            \item YES, if there exists a quantum state $\ket{\psi}\in\mathbb{C}^{2^n}$ such that $\expec{\psi}{H}\le a$.
            \item NO, if $\expec{\psi}{H}\ge b$ for all $\ket{\psi}\in\mathbb{C}^{2^n}$.
        \end{itemize}
    \end{itemize}
    In other words, $k$LH problem is asked to decide whether $\lambda(H)\le a$ or $\lambda(H)\ge b$.
    We say an algorithm $A_{LH}$ solves $kLH(H, a, b)$ if $A_{LH}$ decides $(H,a,b)$ correctly.
\end{definition}

\begin{definition}[Approximating quantum partition function of $k$-local Hamiltonian ($kQPF$) problem]
    \label{def:QPF}
    The quantum partition function problem is to approximate the partition function of a $k$-local Hamiltonian up to a multiplicative error under a certain temperature.
    \begin{itemize}
        \item \textbf{Inputs:} a $k$-local Hamiltonian $H$ acting on $n$ qubits where $k=O(1)$, an inverse temperature $\beta\le1/\poly(n)$, and an error parameter $\delta\in(0,1)$.
        \item \textbf{Outputs:} $\zout\in\mathbb{R}$ such that
        \begin{equation}
            (1-\delta)Z \leq \zout \leq (1+\delta)Z,
        \end{equation}
        where $Z:= \tr{e^{-\beta H}}$.
        \end{itemize}
        We say an algorithm $A_{QPF}$ solves $kQPF(H, \beta, \delta)$ if $A_{QPF}$ outputs such $\zout$ on the inputs $H, \beta, \delta$.
\end{definition}
From now on, when we use the term QPF or $k$QPF, it means to approximate the quantum partition function up to a relative error, instead of to compute the exact value.

\subsection{Satisfiable problem and the quantum strong exponential time hypothesis}
\label{sec:qseth}

Our lower bound for $k$LH and $k$QPF comes from the hardness of satisfiability problem.
Here we formally define the satisfiability problem. 

\begin{definition}[Satisfiability for $k$CNF ($k$SAT) problem]
    \label{def:kSAT}
    The satisfiability problem is a decision problem that asks whether there is an assignment satisfying the given $k$CNF formula.
    \begin{itemize}
        \item \textbf{Inputs:} a $k$CNF formula $\Phi$ that contains $n$ variables. The number of clauses of $\Phi$ is $m=\poly(n)$.
        \item \textbf{Outputs:}
        \begin{itemize}
            \item YES, if there exists an assignment $x\in\{0,1\}^n$ such that $\Phi(x)=1$.
            \item No, if $\Phi(x)=0$ for all $x\in\{0,1\}^n$. 
        \end{itemize}
    \end{itemize}
    We say an algorithm $A_{SAT}$ solves $kSAT(\Phi)$ if $A_{SAT}$ decides $\Phi$ correctly.
\end{definition}

Though the exact lower bound for $k$SAT problem is still unknown, it is widely believed that brute-force search is optimal for classical algorithm and Grover search is optimal for quantum algorithm.
Therefore we assume the lower bound for $k$SAT with $n$ variables is $\Omega(2^\frac{n}{2})$.
The conjecture that to solve $k$SAT needs $\Omega(2^\frac{n}{2})$ is called quantum strong exponential time hypothesis (QSETH).
We formally state QSETH below.
\begin{conjecture}[Quantum strong exponential time hypothesis (QSETH)\cite{ACL+20, BPS21}]
\label{conj:QSETH}
For all $\varepsilon>0$, there exists $k, n_0\in\mathbb{N}$ 
such that for any quantum algorithm $A_{SAT}$, for all $n\ge n_0$ there exists a $k$CNF formula $\Phi$ containing $n$ variables and the number of clauses is $m=O(n^c)$ where $c\in[1,2)$ such that $A_{SAT}$ cannot solve $kSAT(\Phi)$ with probability greater than $\frac{2}{3}$ in $O(2^{\frac{n}{2}(1-\varepsilon)})$.
\end{conjecture}

\subsection{Fine-grained reduction}
\label{sec:fine-grained_reduction}
We use fine-grained reduction to prove the lower bound of $k$LH and $k$QPF problems.
Here we introduce the fine-grained reduction. 
\begin{definition}[Fine-grained reduction \cite{VW15}]
    \label{def:fine-grain_reduction}
    Let $P$ and $Q$ be two problems and $A_Q$ be an oracle that solves $Q$ with probability greater than $\frac{2}{3}$. 
    Let $p(\cdot)$ and $q(\cdot)$ be two non-decreasing functions.  
    We say $P$ is $(p,q)$ reducible to $Q$ if for all $\varepsilon$, there exist $\xi$, an algorithm $A_P$, a constant $d$, and an integer $r(n)$ such that the algorithm $A_P$ that can black-boxly assess to $A_Q$ takes an instance of $P$ with size $n$ and satisfies the following.
    \begin{enumerate}[label=\roman*.]
        \item \label{requre1} $A_P$  solves $P$ with probability greater than $\frac{2}{3}$.
        \item \label{requre2} $A_P$ runs in $d\cdot p(n)^{1-\xi}$ time.
        \item \label{requre3} $A_P$ produces at most $r(n)$ instances of $Q$ adaptively.
        \item \label{requre4} $\sum_{i=1}^{r(n)}(q(n_i))^{1-\varepsilon} \le d\cdot (p(n))^{1-\xi}$, where $n_i$ is the size of the $i$th instance of problem $Q$ that is produced by $A_P$.
    \end{enumerate}
         
\end{definition}
One can also use a more general reduction, \emph{quantum fine-grained reduction}~\cite{ACL+20}.
In a quantum fine-grained reduction, the reduction algorithm $A_P$ is allowed to query the oracle $A_Q$ in superposition.
In this work, however, the definition provided in \Cref{def:fine-grain_reduction} is sufficient to establish the lower bounds for both the Local Hamiltonian problem and the approximation of the Quantum Partition Function.

When $P$ is $(p,q)$ reducible to $Q$ and every instance produced by $A_P$ has size $n+o(n)$, we have that if $P$ cannot be solved within time $O(p(n)^{1-\varepsilon})$ for any $\varepsilon$, then $Q$ cannot be solved within time $O(q(n)^{1-\xi})$ for any $\xi$.

Now we are going to show that $k$LH reduces to $k$QPF through a fined-grained reduction.

\begin{lemma}[Fine-grained reduction from $k$LH to $k$QPF]
    \label{lem:FG_reduction_of_QPF}
    Let $T(n)\in\omega(\poly(n))$. $k$LH($H,a,b$) is $(T(n), T(n))$ reducible to $k$QPF($H,\beta, \delta$) in which $H$ acts on $n$ qubits, $\beta\geq \frac{n}{b-a}$ and $\delta$ satisfies that $\frac{1-\delta}{1+\delta}\geq e^{-0.3n}$.
\end{lemma}
We emphasize that in Lemma~\ref{lem:FG_reduction_of_QPF},  $H$ in the $k$QPF problem and the $k$LH problem is the same Hamiltonian.

    To better understand Lemma~\ref{lem:FG_reduction_of_QPF}, we unpack the underlying fine-grained reduction as follows. 
    Suppose there exist $\xi>0$ and an algorithm $A_{QPF}$ that approximate $\tr{e^{-\beta H}}$ up to relative error $\delta$ satisfying $\frac{1-\delta}{1+\delta}\ge e^{-0.3n}$ with probability greater than $2/3$ in $O(T(n)^{1-\xi})$ time, 
    then for any $\varepsilon>0$, we can use $A_{QPF}$ to construct an algorithm $A_{LH}$ such that given thresholds $a,b$ satisfying $b-a\ge n/\beta$, the algorithm $A_{LH}$ decides weather $\lambda(H)\ge b$ or $\lambda(H)\le a$ with probability greater than $2/3$ in $O(T(n)^{1-\varepsilon})$ time.  

    On the other hand,
    suppose for all $\varepsilon>0$, for any algorithm $A_{LH}$, for infinitely many $n$ there exists a $k$-local Hamiltonian $H_n$ acting on $n$ qubits and two energy thresholds $a,b$ such that $A_{LH}$ cannot solve $kLH(H_n,a,b)$ with probability greater than $\frac{2}{3}$ in $O(T(n)^{1-\varepsilon})$ time,    
    then for all $\xi>0$, for all $\delta>0$, for infinitely many $n\ge n_1$, 
    any algorithm $A_{QPF}$ cannot solve $kQPF(H_n, \beta, \delta)$, where $\beta\ge\frac{n}{b-a}$, with probability greater than $2/3$ in $O(T(n)^{1-\xi})$ time.
    The integer $n_1$ satisfies that $\frac{1-\delta}{1+\delta}=e^{-0.3n_1}$.

\begin{proof}[Proof of Lemma~\ref{lem:FG_reduction_of_QPF}]
    Let $(H,a,b)$ be the $k$LH instance, where $H$ is a $k$-local Hamiltonian acts on $n$ qubits .
    We are going to construct an algorithm $A_{LH}$ that solves $kLH(H, a, b)$ by using $A_{QPF}$ as a subroutine.
    
    If $(H, a, b)$ is YES case, then there is at least one eigenstate of $H$ whose energy is lower than or equal to $a$.
    Hence $Z(\beta)\ge e^{-\beta a}$.
    
    If $(H, a, b)$ is NO case, then all $2^n$ number of eigenstates of $H$ have energy higher than or equal to $b$.
    Hence, $Z(\beta)\le 2^n e^{-\beta b} < e^{-\beta b +0.7n}$.
    
    When $\beta \ge \frac{n}{b-a} \in \poly(n)$, and $\delta$ satisfies $\frac{1-\delta}{1+\delta} \ge e^{-0.3n}$, it holds that 
    \[
    \frac{1-\delta}{1+\delta}\ge e^{-0.3n}\ge e^{-\beta(b-a)+0.7n}.
    \] 
    Hence we have $(1-\delta)e^{-\beta a}\ge (1+\delta)e^{-\beta b +0.7n}$.
    
    Now we present the algorithm $A_{LH}$ that solves $kLH(H, a, b)$ by using $A_{QPF}$.
    \begin{enumerate}
        \item When receiving $H, a, b$, calculate $\delta_0$ such that $\frac{1-\delta_0}{1+\delta_0} = e^{-0.3n}$ and $\beta_0=\frac{n}{b-a}$.
        Set $\beta \ge\beta_0$ and set $\delta \le\delta_0$.
        \item Run $A_{QPF}$ on the input $H, \beta, \delta$, and get the output $\zout$.
        \item If $\zout\ge (1-\delta)e^{-\beta a}$, then output YES.
        If $\zout\le (1+\delta)e^{-\beta b +0.7n}$, then output NO.
    \end{enumerate}
    When $A_{QPF}$ solves $kQPF(H, \beta, \delta)$ successfully, it is guaranteed that $(1-\delta)Z(\beta) \le \zout \le (1+\delta)Z(\beta)$.
    When the inputs of $k$LH is YES case, $\zout\ge (1-\delta)Z(\beta) \ge (1-\delta)e^{-\beta a}$; and when NO case, $\zout\le (1+\delta)Z(\beta) < (1+\delta)e^{-\beta b +0.7n}$.

    By the choice of $\beta$ and $\delta$, it holds that $(1-\delta)e^{-\beta a}\ge (1+\delta)e^{-\beta b +0.7n}$.
    Therefore, when $A_{QPF}$ solves $kQPF(H, \beta, \delta)$ successfully, $A_{LH}$ decides $kLH(H, a, b)$ correctly.
    The success probability of $A_{LH}$ is the same as $A_{QPF}$.
    Therefore, the requirement \ref{requre1} in Definition \ref{def:fine-grain_reduction} is satisfied.

    The algorithm $A_{LF}$ queries $A_{QPF}$ once in Step 2., and the instance of $k$QPF is exactly the input of $k$LH.
    For any $\varepsilon$, we choose $\xi=\varepsilon$. 
    We have $T(n)^{1-\varepsilon}\le d\cdot T(n)^{1-\xi}$ for any constant $d>1$.
    Therefore, the requirement \ref{requre3} and \ref{requre4} in Definition \ref{def:fine-grain_reduction} are satisfied.

    Finally, Step 1.\ and Step 3.\ in the algorithm $A_{LH}$ run in $\poly(n)$ time.
    Hence, $A_{LH}$ runs in $\poly(n)$ times, which is less than $d\cdot T(n)^{1-\xi}$ for some constant $d$ because $T(n)$ is superpolynomial.
    consequently,  the requirement \ref{requre2} is satisfied.
    This finishes the proof.
\end{proof}

%% file: 4_0_lower_bound.tex
\section{Lower bound for $k$-local Hamiltonian}
We first present lower bounds for the 5-local LH and QPF problems in \Cref{sec:5local}, and then present the lower bounds for 3-local cases in \Cref{sec:3local}.

\input{4_1_0_five_local}

\input{4_2_three_local}

%% file: 4_1_0_five_local.tex
\subsection{Lower bound for 5-local Hamiltonian}\label{sec:5local}
We present the lower bound for the 5LH problem in the following theorem. 

\begin{theorem}[Lower bound for 5LH]
    \label{thm:main_lh}
    Assume QSETH holds.
    Then, for any $\xi>0$, for any quantum algorithm $A_{LH}$, for infinitely many $n_H$, there exists a $5$-local Hamiltonian $H$ acting on $n_H$ qubits,
    associated with $a,b$ satisfying $b-a = \Delta(n_H)$ where $O(1/n_H^6)<\Delta(n_H)<O(1/n_H^3)$, such that $A_{LH}(H, a, b)$ cannot decide $5LH(H, a, b)$ with probability greater than $2/3$ in $O(2^{\frac{n_H}{2}(1-\xi)})$ time.
\end{theorem}

We emphasize that the norm of each term in $H$ in \Cref{thm:main_lh} is bounded by 1, as will be shown later in the proof.

The lower bound for the 5QPF problem is described in the following theorem.

\begin{theorem}
    [Lower bound for 5QPF]
    \label{thm:main_qpf}
    Assume QSETH holds.
    For any $\xi>0$ and any $\delta\in(0,1)$, 
    for any quantum algorithm $A_{QPF}$, for infinitely many $n_H$,
    there exists a 5-local Hamiltonian $H$ acting on $n_H$ qubits, associated with an inverse temperature $\beta_0=O(n_H^7)$, such that for all $\beta\ge \beta_0$, the algorithm $A_{QPF}$ cannot solve $5QPF(H,\beta, \delta)$ in $O(2^{\frac{n_H}{2}(1-\xi)})$ time with probability greater than $2/3$.
\end{theorem}

\begin{proof}
    We prove the theorem by contradiction.
    Assume that there exists an algorithm $A_{QPF}$ that solves the 5QPF problem.
    Then, by Lemma~\ref{lem:FG_reduction_of_QPF}, we can construct an algorithm $A_{LH}$ that solves 5LH, which contradicts \Cref{thm:main_lh}.  
     
    Suppose there exist $\xi>0$ and $\delta\in(0,1)$ such that there exist an algorithm $A_{QPF}$ and $n_1\in\mathbb{N}$ satisfying the following: for all $n_H>n_1$, for all $H$ acting on $n_H$ qubits, and for arbitrarily large $\beta\ge O(n_H^7)$, the algorithm $A_{QPF}$ solves $kQPF(H,\beta, \delta)$ with probability greater than $2/3$ in $O(2^{\frac{n}{2}(1-\xi)})$ time.
    
    Let $(H,a,b)$ be an LH instance, where $H$ is a $k$-local Hamiltonian acting on $n_H$ qubits with sufficiently large $n_H$ (we will specify how large $n_H$ shall be later), and $b-a=\Delta(n_H)$ that satisfies $O(1/n_H^6)<\Delta(n_H)<O(1/n^3)$.
    Choose $\beta$ such that $\beta \ge O(n^7)>\frac{n}{b-a}$.
    By Lemma~\ref{lem:FG_reduction_of_QPF}, we construct $A_{LH}$, in which we query $A_{QPF}$ on the instance $(H, \beta, \delta)$.
    The algorithm runs in $O(2^{\frac{n}{2}(1-\xi)})$ time.
    
    Let $n_2$ be the smallest integer such that $\frac{1-\delta}{1+\delta}\ge e^{-0.3n_2}$.
    By the hypothesis at the beginning of the proof and by Lemma~\ref{lem:FG_reduction_of_QPF}, when $n_H\ge\max\{n_1,n_2\}$, the algorithm $A_{LH}$ decides $(H,a,b)$ successfully with probability greater than $2/3$.

    Note that the reduction works for all $H$ acting on $n_H$ qubits and $a,b$ satisfying $O(\frac1/n_H^3)<b-a<O(1/n_H^6)$ as long as $n\ge\max\{n_1,n_2\}$.
    We thus obtain a contradiction with \Cref{thm:main_lh}.
    This completes the proof.
\end{proof}

To prove \Cref{thm:main_lh}, we will use the following lemmas.

\begin{lemma}[A quantum circuit can compute $k$CNF]
    \label{lem:U_compute_Phi}
    For any positive integer $k$ and $c>0$, for all $n\in\mathbb{N}$, for any $k$CNF formula $\Phi$ that contains $n$ variables and $m=O(n^c)$ clauses, there exists a quantum circuit $U_\Phi$ that acts on $n$ input qubits and at most $2c\log n +2$ ancilla qubits, and $U_\Phi$ consists of at most $34c^2 n^c \log^2 n + (70k+2)n^c + 35c\log n$ elementary gates such that $U_{\Phi}\ket{x}\reg{in}\otimes\ket{0}\reg{anc}=\ket{\Phi(x)}\reg{out}\otimes\ket{\psi_x}\reg{\overline{out}}$ for all $x\in\{0,1\}^n$, where $\ket{\psi_x}$ is some quantum state depending on $x$.
    The construction of $U_{\Phi}$ can be done in $\poly(n)$ time.
\end{lemma}

We defer the proof to \Cref{sec:ksatcirc}.

\begin{lemma}[Circuit-to-Hamiltonian reduction]
    \label{lem:YaoTingH}
    For all $n\in\mathbb{N}$, 
    for quantum circuit $U$, whose number of input qubits is $n$, number of ancilla qubits is $n_a(n)$, and number of elementary gates is $g(n)$, where $n_a(n)$ and $g(n)$ are arbitrary integers,
    there exist a $(d+3)$-local Hamiltonian $H_U$ acting on $n_H:=n+n_a(n)+\nclock$ qubits, where $d$ is any positive integer and $\nclock$ is a positive integer satisfying $\binom{\nclock}{d} \ge g(n)$, associated with $a,b$ satisfying $b-a = 1/\Delta$ where $\Delta= O(1/\binom{\nclock}{d}^3)$ such that for all $\mu\in\negl(n)$ the following hold:
    \begin{itemize}
        \item If there exists $\ket{\psi}\in\mathbb{C}^{2^n}$ such that $\Pr[1\gets U(\ket{\psi})]\ge 1-\mu$, then $\lambda(H_U)\le a$, and
        \item If $\Pr[1\gets U(\ket{\psi})]\le \mu$ for all $\ket{\psi}\in\mathbb{C}^{2^n}$, then $\lambda(H_U)\ge b$.
    \end{itemize}
    The construction of $H_U$ can be performed in $\poly(n)$ time.
\end{lemma}
We defer the proof to \Cref{sec:c2h}.

\begin{remark}[Size-preserving circuit to Hamiltonian reduction]
    In \Cref{lem:YaoTingH}, we can find a $c$ such that $g= O(n^{c})$, and let $d$ be the smallest integer such that $d>c$. 
    Then, we can choose $\nclock= O(n^{\frac{c}{d}})$ such that $\binom{\nclock}{d} \ge g$.
    We can find such $\nclock$ because $\binom{\nclock}{d} = O(n^{c})$.
    Hence, we have that $\nclock\in o(n)$.
    If $n_a$ is also in $o(n)$, then $n_H=n+o(n)$.
    
    For example, if $g=O(n^{1.5}\log n)$, we can choose $c=1.7$, $d=2$, and $\nclock=n^{0.85}$.
\end{remark}
Now we are ready to prove our main result.
\begin{proof}[Proof of \Cref{thm:main_lh}]
    We reduce $k$SAT to 5LH.
    Suppose there exist $\xi>0$ and an algorithm $A_{LH}$ such that there exists $n_1\in\mathbb{N}$ satysfying the following: 
    for all $H$ acting on $n_H\ge n_1$ qubits and for all $a,b$ satisfying $O(\frac{1}{n_H^6}) < b-a< O(\frac{1}{n_H^3})$, the algorithm $A_{LH}$ solves $(H,a,b)$ with probability greater than $\frac{2}{3}$ in $T_{\xi}(n_H)\in\omega(\poly(n_H))$ times. 
    The running time  $T_{\xi}(n_H)$ will be determined later.
    
    Let $\Phi=\varphi_1\wedge\varphi_2\wedge\cdots\wedge\varphi_m$ be a $k$CNF formula defined on $n$ variables $x_1,x_2,\dots,x_n$, where $m=O(n^c)$ and $c\in[1,2)$. 
    We construct an algorithm $A_{SAT}$ that solves $kSAT(\Phi)$
    by using $A_{LH}$ as a subroutine. 
    
    \begin{enumerate}
        \item Upon receiving a $k$SAT instance $\Phi$, construct a quantum circuit $U_\Phi$ from Lemma~\ref{lem:U_compute_Phi}.
        \item Construct a Hamiltonian $H_{U_\Phi}$ acting on $n_H=n+o(n)$ qubits from $U_\Phi$, and obtain two energy thresholds $a, b$ where $b-a = O(1/\binom{\nclock}{d}^3)$ by Lemma~\ref{lem:YaoTingH}.
        (The details of the choice of $\nclock$ and $d$ will be stated later.)
        \item Run $A_{LH}$ on the input $(H,a,b)$.
         If the output of $A_{LH}(H,a,b)$ is YES, then return YES.
         If the output of $A_{LH}(H,a,b)$ is NO, then return NO.
    \end{enumerate}
    The running time of $A_{SAT}$ is $T_{\xi}(n) + \poly(n)$.

    We  now show the correctness of $A_{SAT}$.
    By Lemma \ref{lem:U_compute_Phi}, we have the following: 
    \begin{itemize}
        \item If there exists an assignment $x\in\{0,1\}^n$ such that $\Phi(x)=1$, then $\Pr[1\gets U_\Phi(\ket{x})]\ge 1-2^{-n}$.
        \item If $\Phi(x)\neq 1$ for all assignments $x\in\{0,1\}^n$, then $\Pr[1\gets U(\ket{x})]\le 2^{-n}$ for all $x\in\{0,1\}^n$. 
    \end{itemize}
    Combining Lemma~\ref{lem:YaoTingH}, we obtain:
    \begin{itemize}
        \item If there exists $x\in\{0,1\}^n$ such that $\Phi(x)=1$, then $\lambda(H_{U_\Phi})\le a$.
        \item If $\Phi(x)\neq 1$ for all $x\in\{0,1\}^n$, then $\lambda(H_{U_\Phi})\ge b$. 
    \end{itemize}
    Therefore, $A_{\mathrm{SAT}}$ has the same success probability as $A_{\mathrm{LH}}$ 

    Next, we show that $n_H = n + o(n)$ and that $H_{U_{\Phi}}$ is 5-local.
    Let $n_a$ and $g$ denote the number of ancilla qubits and elementary gates in $U_{\Phi}$, respectively.
    By Lemma~\ref{lem:U_compute_Phi}, $n_a=o(n)$, and $g\le 34 c^2 n^c \log^2 n + (70k + 2)n^c + 35c \log n$. 
    There exists $n_2\in\mathbb{N}$ such that for all $n \ge n_2$, the gate number $g$ is upper-bounded by $35 c^2 n^{c'}$, where $c' \in (c, 2)$ is a constant. 
     
    By Lemma~\ref{lem:YaoTingH},   
    the number of qubits of $H_{U_\Phi}$ is $n_H=n+n_a+\nclock$, where $\nclock$ satisfies $\binom{\nclock}{d}\ge g$, 
    and $H_{U_\Phi}$ is $d+3$ local.
    
    Since $c' < 2$, we choose $d = 2$ and set $\nclock = r n^{c'/d}$, where $r$ is a constant such that $\binom{\nclock}{d} \ge 35 c^2 n^{c'} \ge g$ for all $n \ge n_2$. 
Because $\binom{\nclock}{d} = O(n^{c'})$, there exists $r = O(1)$ such that $\binom{\nclock}{d} \ge 35 c^2 n^{c'}$. 
It follows that $\nclock \in o(n)$. 
As a result, the total number of qubits is $n + n_a + \nclock = n + o(n)$. 
Moreover, $H_{U_{\Phi}}$ is 5-local.
Furthermore, we have that the threshold gap $\Delta = 1/\binom{\nclock}{d}^3 = O(1/n^{3c'})$, which satisfies $O(1/n_H^6)<\Delta(n_H)<O(1/n_H^3)$.

    Finally, we show this reduction contradicts QSETH.
    Because $n_H=n+o(n)$, we have $n_H\le (1+\eta)n$ for all $n\ge n_3$ for some constant and $\eta<1$ and $n_3$.

    Set $\varepsilon=\xi-\eta+\xi\eta$ and $T_{\xi}(n_H) = O(2^{\frac{n_H}{2}(1-\xi)})$.
    When $n \ge \max\{n_1, n_2, n_3\}$, the algorithm $A_{SAT}$ decides $kSAT(\Phi)$ within running time 
    $O(2^{\frac{n_H}{2}(1-\xi)})\le O(2^{\frac{n}{2}(1+\eta)(1-\xi)})= O(2^{\frac{n}{2}(1-\varepsilon)})$ time. 

    Note that the reduction works for any $k$CNF formula $\Phi$ defined on $n$ variables with $m=O(n^c)$ clauses as long as $n\ge\{n_1,n_2,n_3\}$.
    This leads to a contradiction with QSETH (Conjecture~\ref{conj:QSETH}).
\end{proof}

\input{4_1_1_SAT2C}
\input{4_1_2_C2H}

%% file: 4_1_1_SAT2C.tex
\subsubsection{Proof of Lemma \ref{lem:U_compute_Phi}: constructing a circuit that calculates $k$SAT}
\label{sec:ksatcirc}
In this section, we show the construction of the quantum circuit $U_\Phi$ that computes the formula $\Phi=\varphi_1\wedge\dots\wedge\varphi_m$ defined on $n$ variables $x_1,x_2,\dots x_n$, and $m=O(n^c)$.
For each $i\in[m]$, the clause $\varphi_i=\ell_{i,1}\vee \ell_{i,2}\vee\cdots\vee\ell_{i,k_i}$ where for each $p\in[k_i]$, $\ell_{i,p}$ can be either $x_j$ or $\neg x_j$ where $j\in [n]$, and $k_i\le k$.
Let $r:=\ceil{\log m}$.

\begin{proof}[Proof of Lemma \ref{lem:U_compute_Phi}]
    The ideal is to compute $\varphi_i$ for each $i\in[m]$ sequentially.
    We initially set a counter, and increment the counter by one if $\varphi_i(x)=1$.
    After compute all $\varphi_i$, we check weather the counter is equal to $m$.
    Let $k_i$ be the number of variables ($x_j$ or $\neg x_j$) appearing in $\varphi_i$.
    We have $\Phi(x)=1$ if and only if the counter equals $m$.
    We introduce the ancilla register $\rgst{anc}=\rgst{cls}\cup \rgst{cnt} \cup \rgst{out}$ where $\rgst{cls}$ is a one-qubit register that temporarily stores the value of $\varphi_i(x)$ for each $i$, $\rgst{cnt}$ is an $r$-qubit ($r=\ceil{\log m}$) that serves as the counter, and $\rgst{out}$ is a one-qubit register that outputs $\Phi(x)$.

    To compute each $\varphi_i$, we construct $W_i$ acting on $\rgst{in}\cup \rgst{cls}$ such that $W_i\ket{x}\reg{in}\ket{0}\reg{cls}=\ket{\psi_x}\reg{in}\ket{\varphi_i}\reg{cls}$, where $\ket{\psi_x}$ is some state depending on $x$. 

    To increment the counter by one, we construct a unitary $\addone$ acting on $\rgst{cnt}$ such that $\addone\ket{y}\reg{cnt}=\ket{bin(int(y)+1)}\reg{cnt}$ for all $y\in\{0,1\}^r\setminus\{1^r\}$.
    We apply $\addone$ only when $\varphi_i(x)=1$. 
    This is done by letting $\addone$ be controlled by the resister $\rgst{cls}$.
    Denote the control-$\addone$ operator by $C\addone$.
    Then $C\addone\reg{cls\cup cnt}\ket{0}\reg{cls}\ket{y}\reg{cnt}=\ket{0}\reg{cls}\ket{y}\reg{cnt}$ and $C\addone\reg{cls\cup cnt}\ket{1}\reg{cls}\ket{y}\reg{cnt}=\ket{0}\reg{cls}\ket{bin(int(y)+1)}\reg{cnt}$.

    After calculating $\varphi_i(x)$ and applying $C\addone$, we apply $W_i^\dagger$ on $\rgst{in}\cup \rgst{cls}$ to restore the state to $\ket{x}\reg{in}\ket{0}\reg{cls}$. 

    To check whether the counter is equal to $m$ in the final step, we construct a compare operator, denoted by $\compare$, that acts on $\rgst{cnt}\cup\rgst{out}$.
    The operator $\compare$ satisfies $\compare \ket{y}\reg{cnt}\ket{0}\reg{out}=\ket{y}\reg{cnt}\ket{\delta_{m,int(y)}}\reg{out}$ for all $y\in\{0,1\}^{r}$. 

    To sum up, we construct the quantum circuit $U_{\Phi}$ as follows: 
    \begin{equation}
        U_\Phi:= \compare(W_m^\dagger C\addone W_m) (W_{m-1}^\dagger C\addone W_{m-1})\cdots(W_1^\dagger C\addone W_1).
    \end{equation}

    Next, we explain how to implement $W_i$, $\addone$, and $\compare$ gates.

    Because $\ell_{i_1}\vee \ell_{i_2}\vee\cdots\vee \ell_{i_{k_i}} = \neg(\neg \ell_{i_1} \wedge \neg \ell_{i_2}\wedge \cdots \wedge \neg \ell_{i_{k_i}})$, we can use $C^k\notgate$ together with $\notgate$ gates to implement $W_i$.
    Let $\rgst{S_i}\subseteq\rgst{in}$ be the register defined by $\rgst{S_i}:=\{j\in[n]:x_j\in\varphi_i\mathrm{\;or\;}\neg x_j \in\varphi_i \}$, i.e., $\rgst{S_i}$ consists of the $j$th qubits in $\rgst{in}$ such that $x_j$ or $\neg x_j$ in $\varphi_i$;
    and let $\rgst{R_i}\subseteq\rgst{in}$ be the register defined by $\rgst{R_i}:=\{j\in[n]:x_j\in\varphi_i\}$, i.e., $\rgst{R_i}$ consists of the $j$th qubits in $\rgst{in}$ such that $x_j$ in $\varphi_i$, but $\neg x_j$ does not. 
    We construct $W_i$ by 
    $W_i:=\notgate\cdot\reg{cls}C^{k_i}\notgate\reg{S_i\cup cls}\cdot\notgate^{\otimes |\rgst{R_i}|}\reg{R_i}$.

    To implement the $\addone$, observe that for $y\in\{0,1\}^r$, when $int(y)$ is incremented by one, a bit $y[p]$ will be flipped if and only if all the bits with order lower than $p$ are 1's.
    That is, $y[p]$ will be flipped if and only if for all $p'>p$, $y[p'] = 1$.
    Hence, $\addone$ can be composed of a sequence of multi-control-NOT gates.
    The $q$th layer of $\addone$ is a $C^{r-q}\notgate$ gate whose control qubits are ${q,q+1,\dots,r}$ and whose target is the $q$the qubit.
    The last layer of $\addone$ is a $\notgate$ gate acting on the last qubit.
    To implement the control-$\addone$ operation $C\addone$, we let every $C^{r-q}\notgate$ be controlled by register $\rgst{cls}$.
    In other words, the  $q$th layer of $C\addone$ is a $C^{r+1-q}\notgate$ gates whose control qubits are ${q,q+1,\dots,r}\cup\rgst{cls}$ and whose target is the $q$th qubit in $\rgst{cnt}$.

    The $\compare$ operator checks whether the value stored in the counter equal to $m$.
    That is, $\compare$ operator compares each bit in the counter with $bin(m)$, which can be implemented by a $C^r\notgate$ gate. 
    Let $\rgst{P}\subseteq\rgst{cnt}$ be defined by $\rgst{P}:=\{j\in[r]:bin(m)[j]=0\}$.
    We construct $\compare$ by $\compare:=C^r\notgate\reg{cnt\cup out}\cdot \notgate^{\otimes |P|}\reg{P}$.

    We decompose each multi-control-NOT gate into Toffoli gates.
    For $W_i$ (and $W_i^\dagger$), there is a $C^{k_i}\notgate$ inside, and it can be decomposed into at most $2k$ of Toffoli gates using at most $k$ ancilla bits.

    The largest gate in the $C\addone$ is a control $C^r\notgate$ gate, and there are $r$ layers.
    Hence, $C\addone$ can be decomposed into at most $2r^2$ Toffoli gates.
    Because the ancillas for the multi-control-NOT gates can be reused, the ancillas required for $C\addone$ is at most $r$.

    The $\compare$ operation contains a $C^{r}\notgate$ gate, which can be decomposed into at most $2r$ Toffoli gates using at most $r$ ancilla qubits.

    The ancilla qubits can be reused for the multi-control-NOT gates. 
    Hence, to decomposed $U_\Phi$ into Toffoli gates, the number of  ancilla qubits required is at most $r$. 
    Therefore, the total number of ancilla qubits for $U_\Phi$ is $|\rgst{anc}|+r = |\rgst{cls}|+|\rgst{cnt}|+|\rgst{out}|+r = 2\log m +2 = 2c\log n +2$.

    The total number of Toffoli gates in $U_\Phi$ is at most $(4k+2r^2)\cdot m + 2r = 2c^2 n^c \log^2 n + 4k n^c +2c\log n$.
    We further decompose the Toffoli gates into elementary gates,
    yeilding $34c^2 n^c \log^2 n + 68kn^c +34c\log n$ elementary gates in total.
    In addition to Toffoli gates, there are at most $2(k+1)\cdot m + r= (2k+2)n^c +  c\log n$ of $\notgate$ gates.
    Therefore, the total number of gates is at most $34c^2 n^c c^2 \log^2 n + (70k+2)n^c +35c \log n$.
\end{proof}

%% file: 4_1_2_C2H.tex
\subsubsection{Proof of Lemma \ref{lem:YaoTingH}: space-preserving circuit to Hamiltonian reduction}\label{sec:c2h}
In this section, we construct a local Hamiltonian $H_U$ whose ground state energy depends on the output of a quantum circuit $U$ on the input $\ket{\psi}$.

Following the idea proposed by Kitaev et.al.\cite{KSV02}, we introduce a clock register.
The Hamiltonian $H_U$ acting on the circuit register union clock register such that the ground state of $H_U$ encodes the computation process of $U\ket{\psi}\reg{in}\ket{0^{n_a}}\reg{anc}$. 

We use the clock states proposed in \cite{CCH+23} so that the clock register consists of $\nclock=n^{c}$ many of qubits, where $c$ is a constant, that can encode $g=\poly(n)$ steps of computation, while the locality of $H_U$ is a constant.
The constant $c$ can be smaller than 1.
Hence, we can use $o(n)$ qubits for the clock register to encode a $poly(n)$ computation.

The clock states is constructed by a Johnson graph.
The vertices of the Jonson graph are all the subsets of $[n]$ with size $k<n$.
Two vertices are adjacent if and only if the subsets exactly differ by one element. 

\begin{definition}[Johnson graph]
    \label{def:Johnson}
    Let $n,d\in\mathbb{N}$ and $d<n$. We say a Johnson graph $J(n,d)=(V,E)$ is a graph that satisfies the following requirement.
    \begin{itemize}
        \item $V=\{S\subseteq[n]:|S|=k\}$.
        \item $E=\{(S,S')\in V^2:|S\cap S'|=k-1\}$.
    \end{itemize}
\end{definition}
The number of vertices of $J(n,d)$ is $\binom{n}{d}$.
In \cite{Als12}, it has been proved that for all $n,d$, there is a Hamiltonian path\footnote{A Hamiltonian path is a path that visits every vertex in a graph exactly once. Please do not confuse with the physical quantity $H$.} in the Johnson graph $J(n,d)$.
Also, the proof implicitly construct an algorithm that finds the Hamiltonian path in $O(n^d)$ time. 

Now we explain how to construct the clock.
A clock state is a $n$-qubit state.
Each clock state $\ket{\clock{t}}$ corresponds to a vertex $S_t$ in $J(n,d)$.
All the clock states are basis states of the computational basis whose number of 1's equals to $d$.
The $i$th qubit in $\ket{\clock{t}}$ is 1 if and only if $i$ is chosen into $S_t$.
In other words,
when we treat the clock register as the set $[n]$ and let $\rgst{S_t}$ be the portion therein corresponding to $S_t$,
the state in $\rgst{S_t}$ is $\ket{1^d}$.
The next clock $\ket{\clock{t+1}}$ corresponds to the vertex $S_{t+1}$ that is adjacent to $S_{t}$.
To update the $\ket{\clock{t}}$ to $\ket{\clock{t+1}}$, we make $d-1$ of 1's unchanged and flip two bits.
Hence the update of the clock state can be done with constant local operation.
We define the clock state formally as follows.
\begin{definition}[$(n,d)$-clock state \cite{CCH+23}\footnote{The idea that using the Johnson graph encodes the clock state was proposed by Yao-Ting Lin. We would like to call it a ``Yao-Ting clock state''.}]
    A $(n,d)$-clock state is a collection of $n$-qubit quantum state $\{\ket{\gamma_t}\}_{t=0}^T$ defined by a Johnson graph $J(n,d)=(V,E)$ and $T=\binom{n}{d}-1$. 
    Let $S_0, S_1,\dots, S_{T}\in V$ and the sequence $S_0, S_1,\dots, S_{T}\in V$ forms a Hamiltonian path in $J(n,d)$.
    For any $t\in\{0\}\cup [T]$ the state $\ket{\gamma_t}$ is defined by  
    $\ket{\gamma_t}:=\bigotimes_{i\in[n]}\ket{\indicator_{S_t}(i)}$.
\end{definition}

For all $t\in 0\cup [T]$, let $\rgst{S_t}$ be the corresponding register of $S_t$.
And for all $t\in [T]$
We define the operator $F_t$ as following.
\begin{equation}
    \label{eq:forward}
    F_t:=\ketbra{1}{0}\reg{S_{t-1}\setminus S_{t}}\otimes \ketbra{0}{1}\reg{S_{t}\setminus S_{t-1}}\otimes \proj{1^{d-1}}\reg{S_{t}\cap S_{t-1}}.    
\end{equation}
Hence we have 
\begin{equation}
    \label{eq:backward}
    F_t^\dagger:=\ketbra{0}{1}\reg{S_{t-1}\setminus S_{t+1}}\otimes \ketbra{1}{0}\reg{S_{t}\setminus S_{t}}\otimes \proj{1^{d-1}}\reg{S_{t}\cap S_{t-1}}.    
\end{equation}
It holds that
\begin{equation}
    \label{eq:forward_a_step}
    F_t\ket{\gamma_{t'}}=\delta_{t',t-1}\ket{\gamma_{t'+1}},
\end{equation}
and
\begin{equation}
    \label{eq:backward_a_step}
    F_t^{\dagger}\ket{\gamma_{t'}}=\delta_{t',t}\ket{\gamma_{t'-1}}.
\end{equation}
That is, $F_t$ ``forwards'' the clock $\ket{\clock{t-1}}$ one step, and eliminate all the other state $\ket{\clock{t'}}$ where $t'\neq t-1$.
Likewise, $F_t^\dagger$ ``backwards'' the clock $\ket{\clock{t}}$ one step, and eliminate all the other state $\ket{\clock{t'}}$ where $t'\neq t$.
We have that $F_t$ is $d+1$ local.
Note that $F_t$ is neither unitary nor Hermitian.

For $t\in\{0\}\cup [T]$, we define the operator $P_t$ as following.
\begin{equation}
    \label{eq:pause}
    P_t:= \proj{1^d}\reg{S_t}.
\end{equation}
We have $P_t^\dagger = P_t$.
It holds that
\begin{equation}
    \label{eq:pause_a_step}
    P_t\ket{\gamma_{t'}}=\delta_{t',t}\ket{\gamma_{t'}}.
\end{equation}
That is, $P_t$ ``pauses'' the clock $\ket{\clock{t}}$, and eliminate all the other state $\ket{\clock{t'}}$ where $t'\neq t$.
The operator $P_t$ is $d$ local.

%
%
Now we are ready to prove \Cref{lem:YaoTingH}.
\begin{proof}[Proof of \Cref{lem:YaoTingH}]
    Let $U=V_gV_{g-1}\cdots V_1$ acting on $\rgst{in}\cup \rgst{anc}$.
    Choose $\nclock$ and $d$ such that $\binom{\nclock}{d}\ge g$, and let $T:=\binom{\nclock}{d}-1$.
    Let $\{\ket{\clock{t}}\}_{t=0}^{T}$ be the $(\nclock, d)$-clock.
    
    We define the Hamiltonian $H_U$ that acts on the register $\rgst{in}\cup\rgst{anc}\cup\rgst{clock}$ where $|\rgst{clock}|=\nclock$.
    
    The Hamiltonian $H_U$ is defined as follows.
    \begin{equation}\label{eq:C2H}
        H_{U}:= H_{in}+H_{out} + H_{prop} + H_{stab},
    \end{equation}
    where
    \begin{align}
        H_{in} &:= \sum_i^{n_a}\proj{1}\reg{anc[i]}\otimes\proj{1^d}\reg{S_0},\label{eq:H_in}\\
        H_{out} &:= \proj{0}\reg{out}\otimes\proj{1^d}\reg{S_T},\label{eq:H_out}\\
        H_{prop} &:= \sum_{i=t}^{T}\frac{1}{2}\big(-V_t\otimes F_t -V_t^\dagger \otimes F^\dagger_t + I\otimes P_t + I\otimes P_{t-1}\big),\label{eq:H_prop}
    \end{align}
    where $\rgst{S_0}, \rgst{S_T}\subseteq\rgst{clock}$, 
    $F_t, F_t^\dagger, P_t$ and $P_{t-1}$ act on $\rgst{clock}$, $V_t$ and $V_t^\dagger$ act on $\rgst{in}\cup \rgst{anc}$, and for $t=0$ and $t>g$, $V_t=I$. In addition, $F_t$ is defined in \Cref{eq:forward} and $P_t$ is defined in \Cref{eq:pause}.
    And
    \begin{equation}\label{eq:H_stab}
        H_{stab} :=  H_{>d} + H_{<d} - \frac{\left(\binom{\nclock}{c} - 1\right)}{\binom{\nclock}{c}} I,
    \end{equation}
    where
    \begin{align}
        H_{>d} &:=\sum_{\rgst{S}:|\rgst{S}|=d+1} \ketbra{1^{d+1}}{1^{d+1}}\reg{S},\label{eq:H_more}\\
        H_{<d} &:=\frac{1}{\binom{\nclock}{d}}\sum_{\rgst{S}:|\rgst{S}|=d}\sum_{x\in\{0,1\}^{d}\setminus\{1^d\}} \proj{x}\reg{S},\label{eq:H_fewer}
    \end{align}
    Where $H_{>d}$ and $H_{<d}$ act on $\rgst{clock}$.
    We have that all terms in $H_{in}$ and $H_{out}$ are $(d+1)$-local.
    Since $V_t$ is at most 2-local, $F_t$ is $d+1$ local, and $P_t$ is $d$ local, we have that each term in $H_{prop}$ is $(d+3)$-local.
    In addition, $H_{>d}$ is $d+1$ local, and $H_{<d}$ is $d$ local.
    Hence, $H_U$ is $(d+3)$-local.
    
    The purpose of $H_{stab}$ is to ``give penalty'' to the state whose content in $\rgst{clock}$ is not a clock state.
    Because $H_{stab}$ non-trivially acts on $\rgst{clock}$, we only consider the state defined in $\rgst{clock}$.
    We will see that if $\ket{\phi}$ is a clock state, then $\expec{\phi}{H_{stab}}=0$; otherwise, the energy is high.
    
    The term $H_{>d}$ gives penalty to the state that contains ``too many'' 1's, and the term $H_{<d}$ gives penalty to the state that contains ``too few'' 1's.

    Since $H_{stab}$ is diagonalized in the computational basis i.e., each computational state is an eigenstate of $H_{stab}$. To verify the above statement, we can check $\expec{w}{H_{stab}}$ for all $w\in\{0,1\}^{\nclock}$.

    We divide the Hilbert space of $\rgst{clock}$ into three subspaces $\hs{L}_{=d}$, $\hs{L}_{>d}$, and $\hs{L}_{<d}$ which are defined below.
    \begin{align}
        \hs{L}_{=d} &: =\spn(\{w\in\{0,1\}^{n_a}:wt(w)=d\}),\\
        \hs{L}_{>d} &: =\spn(\{y\in\{0,1\}^{n_a}:wt(y)>d\}),\\
        \hs{L}_{<d} &: =\spn(\{z\in\{0,1\}^{n_a}:wt(z)<d\}).
    \end{align}
    We have that $\hs{L}_{=d}$ is the subspace spanned by the clock states.

    We first calculate $H_{>b}$ acting on the states in $\hs{L}_{=d}$, $\hs{L}_{>d}$, and $\hs{L}_{<d}$ respectively.
    For any $\rgst{S}$ with size $d+1$, $\proj{1^{d+1}}\reg{S}\ket{w} = \ket{w}$ if the subset-string $x_S=1^{c+1}$.
    Hence, we have the follows.
    \begin{itemize}
        \item For all $z\in\{0,1\}^{\nclock}$ such that $wt(z)<d$, it holds that $H_{>d}\ket{z} = 0$ 
        because for all $S\subseteq[\nclock]$ with size $d+1$, $z_S$ has at most $d-1$ of 1's.
        \item For all $w\in\{0,1\}^{\nclock}$ such that $wt(w)=d$, it holds that $H_{>d}\ket{w} = 0$ because for all $S\subseteq[\nclock]$ with size $d+1$, $w_S$ has at most $d$ of 1's.
        \item For all $y\in\{0,1\}^{\nclock}$ such that $wt(y)>d$, it holds that $H_{>d}\ket{y} = r\ket{y}$ where $r\ge 1$, because there is at least one $S\subseteq [\nclock]$ with size $k+1$ such that $y_S=1^{d+1}$. 
    \end{itemize}
    
    Then we calculate $H_{<b}$ acting on the states in $\hs{L}_{=d}$, $\hs{L}_{>d}$, and $\hs{L}_{<d}$ respectively.

    Let $z\in \{0,1\}^{\nclock}$ such that $wt(z)<d$. 
    For any $\rgst{S}$ with size $d$, we have $\sum_{x\in\{0,1\}^d\setminus\{1^d\}}\proj{x}\reg{S}\ket{z} = \ket{z}$, because there exist exactly one $x\in\{0,1\}^d\setminus\{1^d\}$ such that $z_{S}=x$.
    Hence, \[\sum_{\rgst{S}:|\rgst{S}|}\sum_{x\in\{0,1\}^d\setminus\{1^d\}}\proj{x}\reg{S}\ket{z} = \binom{\nclock}{d} \ket{z}.\]

    Let $w\in\{0,1\}^{\nclock}$ such that $wt(w)=d$. 
    Let $S'\subseteq[\nclock]$ be the subset with size $d$ such that $w_{S'}=1^d$.
    For any $\rgst{S}$ with size $d$ such that $S\neq S'$, we have that $\sum_{x\in\{0,1\}^d\setminus\{1^d\}}\proj{x}\reg{S}\ket{w} = \ket{w}$, and
    for $S = S'$, we have $\sum_{x}\proj{x}\reg{S}\ket{w} = 0$;
    Hence, \[\sum_{\rgst{S}:|\rgst{S}|}\sum_{x\in\{0,1\}^d\setminus\{1^d\}}\proj{x}\reg{S}\ket{w} = \left(\binom{\nclock}{d} -1 \right)\ket{w}.\]

    For $y\in\{0,1\}^{\nclock}$ such that $wt(y)>d$,
    since the penalty has been given by $H_{>d}$ already,
    $H_{<d}\ket{y} = p\ket{y}$ where $p\ge 0$ is sufficient for us.
    
    Therefore, we have the follows.
    \begin{itemize}
        \item For all $z\in\{0,1\}^{\nclock}$ such that $wt(z)<d$, it holds that $H_{<d}\ket{z} = \ket{z}$.
        \item For all $w\in\{0,1\}^{\nclock}$ such that $wt(w)=d$, it holds that $H_{<d}\ket{w} = \frac{\binom{\nclock}{d}-1}{\binom{\nclock}{d}} \ket{w}$.
        \item For all $y\in\{0,1\}^{\nclock}$ such that $wt(y)>d$, it holds that $H_{<d}\ket{y} =p\ket{y}$ where $p\ge 0$.
    \end{itemize}
    
    As a result, we have the follows.
    \begin{itemize}
        \item For all $z\in\{0,1\}^{\nclock}$ such that $wt(z)<d$, it holds that $H_{stab}\ket{z} = \frac{1}{\binom{\nclock}{d}}\ket{z} = \frac{1}{1+T}\ket{z}$.
        \item For all $w\in\{0,1\}^{\nclock}$ such that $wt(w)=d$, it holds that $H_{stab}\ket{w} = 0$.
        \item For all $y\in\{0,1\}^{\nclock}$ such that $wt(y)>d$, it holds that $H_{stab}\ket{y} \ge \frac{1}{\binom{\nclock}{d}} \ket{y} = \frac{1}{T+1}\ket{y}$.
    \end{itemize}
    
    Consequently, if $\ket{\phi}\in\hs{L}_{=d}$, then 
    $\expec{\phi}{H_{stab}}=0$, and if $\ket{\phi}\in\hs{L}_{=d}^\perp$, then $\expec{\phi}{H_{stab}}\ge 1/(T+1)$.

    For now, we have shown that the state that is not a clock state has high energy.
    Next, we are going to analyze the state restricted to the clock state. 
    Let 
    \begin{equation}\label{eq:H'}
        H' = H'_{in}+ H'_{out} + H'_{prop},
    \end{equation}  
    where
    \begin{alignat}{2}
        \label{eq:H'_in}
        &H'_{in} &&= (I-\proj{0^{n_a}})\reg{anc}\otimes\proj{\clock{0}}\reg{clock},\\
        \label{eq:H'_out}
        &H_{out} &&= \quad\proj{0}\reg{out}\otimes\proj{\clock{T}}\reg{clock},\\
        \nonumber
        &H'_{prop} &&= \sum_{t=1}^{T}  \frac{1}{2} \Big( -V_t\otimes \ketbra{\clock{t}}{\clock{t-1}}\reg{clock} - V_t^\dagger\otimes \ketbra{\clock{t-1}}{\clock{t}}\reg{clock}\\
        \label{eq:H'_prop}
        &&&\hspace{5em} + I\otimes \proj{\clock{t-1}}\reg{clock} + I\otimes \proj{\clock{t}}\reg{clock}\Big),
    \end{alignat}
    where $V_t$ and $V_t^\dagger$ act on $\rgst{in}\cup\rgst{anc}$.
    
    Let $\hs{S}:=\spn(\{\ket{x}\reg{in\cup anc}\ket{w}\reg{clock}:x\in\{0,1\}^{n+n_a}, w\in\hs{L}_{=d}\})$.
    We have that $\hs{S}^\perp=\spn(\{\ket{x}\reg{in\cup anc}\ket{v}\reg{clock}:x\in\{0,1\}^{n+n_a}, v\in\hs{L}_{=d}^\perp\})$.
    By the construction of the $(n,d)$-clock state, $\proj{1^d}\reg{S_t}\ket{\clock{t'}}=\delta_{t,t'}$.
    Together with the property of $F_t$, $F^\dagger$, and $P_t$, that is, \Cref{eq:forward_a_step}, \Cref{eq:backward_a_step}, and \Cref{eq:pause},
    it holds that $H'\ket{\phi}=H_U\ket{\psi}$
    if $\ket{\phi}\in\hs{S}$. .

    Define an isomorphism $hist$ that maps the Hilbert space of $\rgst{in}\cup\rgst{anc}$ to $\hs{S}$ as follows.
    \begin{equation}\label{eq:hist}
        \ket{hist(\psi)}:= \sum_{t=0}^{T} V_t V_{t-1}\cdots V_{0}\ket{\psi}\reg{in}\ket{0^{n_a}}\reg{anc}\otimes\ket{\gamma_t}\reg{clock}.    
    \end{equation}
    By the result of Kitaev et.al.\cite{KSV02,GHLS15}, we have that

    \begin{itemize}
        \item If there is a quantum state such that $\Pr[1\gets U(\ket{\psi})]\ge 1-\mu$, then $\expec{hist(\psi)}{H'} \le \frac{\mu}{T+1}$.
        \item If for all quantum states $\ket{\psi}$ such that $\Pr[1\gets U(\ket{\psi})]\le \mu$, then $\lambda(H') \ge O(\frac{1-\sqrt{\mu}}{(T+1)^3})$.
    \end{itemize}

    From previous discussion, we have that $H_{stab}\ket{hist(\psi)}=0$.
    Hence $\expec{hist(\psi)}{H_U} \le \frac{\mu}{T+1}$.
    Also, we have that for all $\ket{\phi}\in\hs{S}^\perp$, the energy $\expec{\phi}{H_{stab}} \ge \frac{1}{T+1}$.
    Hence when $\mu=\negl(n)$, the state outside $\hs{S}$ cannot have an energy lower than $\expec{hist(\psi)}{H_U}=\negl(n)$.
    
    Therefore, when $\mu=\negl$,  we have the follows. 
    \begin{itemize}
        \item If there is a quantum state $\ket{\psi}$ such that $\Pr[1\gets U(\ket{\psi})]\ge 1-\mu$, then $\expec{hist(\psi)}{H} \le \frac{\mu}{T+1}$.
        \item If for all quantum states $\ket{\psi}$ such that $\Pr[1\gets U(\ket{\psi})]\le \mu$, then $\lambda(H_U) \ge O(\frac{1-\sqrt{\mu}}{(T+1)^3})$.
    \end{itemize}
    We choose the two energy thresholds $a=\negl(n)$ and $b=O({1}/{T^3})$, we have that the gap $b-a\in O({1}/{T^3})=O({1}/\binom{\nclock}{d}^3)$.
\end{proof}

%% file: 4_2_three_local.tex
\subsection{Lower bound for 3-local Hamiltonian}
\label{sec:3local}
In the previous section, we encode the computation process of a quantum circuit $U$ into a Hamiltonian $H$ by using a $(\nclock, d)$-clock. 
We needs $d+1$-local operators to ``forward'', ``backward'', and ``pause'' the clock state, and we apply 1-local or 2-local operators on the circuit register.
This is the reason that $H$ is $d+3$-local.
When we choose $d=2$, we get a 5-local Hamiltonian.
We can further reduce the locality to 3-local by the technique in \cite{KKR04}.
We present the lower bound for 3LH in the following theorem.

\begin{theorem}[Lower bound for 3LH]
    \label{thm:main:3LH}
    Assume QSETH holds.
    Then, for any $\xi>0$, for any quantum algorithm $A_{LH}$, for infinitely many $n_H$ there exists a $3$-local Hamiltonian $H$ acting on $n_H$ qubits,
    associated with $a,b$ satisfying $b-a = O(1)$ such that $A_{LH}(H, a, b)$ cannot decide $3LH(H, a, b)$ with probability greater than $2/3$ in $O(2^{\frac{n}{2}(1-\xi)})$ time.
\end{theorem}

Note that unlike the Hamiltonian in the previous section, each local term in $H$ in \Cref{thm:main:3LH} is upper-bounded by $\poly(n)$ instead of 1.

By adapting the proof of \Cref{thm:main_qpf}--- replacing \Cref{thm:main_lh} 
with \Cref{thm:main:3LH}---we obtain the lower bound for the 3QPF problem.

\begin{theorem}[Lower bound for 3QPF]
    \label{thm:main:3QPF}
    Assume QSETH holds.
    For any $\xi>0$, and any $\delta\in(0,1)$,  
    for any quantum algorithm $A_{QPF}$, for infinitely many $n_H$, there exists a 3-local Hamiltonian $H$ acting on $n_H$ qubits, associated with an inverse temperature $\beta_0=O(n)$, such that for all $\beta\ge \beta_0$ the algorithm $A_{QPF}$ cannot solves $3QPF(H,\beta, \delta)$ in $O(2^{\frac{n}{2}(1-\xi)})$ time with probability greater than ${2}/{3}$.
\end{theorem}

\Cref{thm:main:3LH} can be proved by modifying the proof of \Cref{thm:main_lh}, where Lemma~\ref{lem:YaoTingH} is replaced with the following Lemma~\ref{lem:3local_main}.

\begin{lemma}[Size-preserving circuit to 3-local Hamiltonian reduction]
    \label{lem:3local_main}
    For any quantum circuit $U$ acting on $n$ input qubits associated with $n_a$ ancilla qubits and $U$ consists of $g=o(n^2)$ gates, there exists a 3-local Hamiltonian acting on $n+n_a+\nclock$ qubits, where $\nclock$ satisfies $\binom{\nclock}{2} \ge 5g$ such that for all $\varepsilon\in(0,1)$, the following two hold.
    \begin{itemize}
        \item If there exists $\ket{\psi}\in\mathbb{C}^{2^n}$ such that $\Pr[1\gets U(\ket{\psi})]\ge 1-\varepsilon$, then $\lambda(H_U)\le \varepsilon$.
        \item If $\Pr[1\gets U(\ket{\psi})]\le \varepsilon$ for all $\ket{\psi}\in\mathbb{C}^{2^n}$, then $\lambda(H_U)\ge \frac{1}{2}-\varepsilon$.
    \end{itemize}
\end{lemma}

One of the key idea in \cite{KKR04} is the projection lemma.

\begin{lemma}[Projection lemma (Lemma 1 in \cite{KKR04})]
    \label{lem:projection_lemma}
    Let $H=H_1+H_2$ be a Hamiltonian acting on the Hilbert space $\hs{S}+\hs{S}^\perp$, where $\|H_1\|\ge 0$ and $\|H_2\|\ge 0$, and $\hs{S}$ is the zero eigenspace of $H_1$.
    Let $J$ be the smallest non-zero eigenvalue of $H_2$,
    If $J>2\|H_1\|$, it holds that 
    \begin{equation}
        \lambda(H)\ge\lambda(H_1\vert_{\hs{S}})-\frac{\|H_1\|^2}{J-2\|H_1\|}.
    \end{equation}
    The notation $H_1\vert_{\hs{S}}$ denotes that $H_1$ is restricted in the subspace $\hs{S}$.
    That is, $H_1\vert_{\hs{S}}=\Pi_{\hs{S}}H_1\Pi_{\hs{S}}$, where $\Pi_{\hs{S}}$ is the projector that projects quantum states on $\hs{S}$.
\end{lemma}

The following lemma is proved in \cite{KKR04}.
\begin{lemma}[Lemma 3 in \cite{KKR04}]
    \label{lem:kkr}
    Let $U=V_TV_{T-1}\cdots V_1$ be a quantum circuit acting on the register $\rgst{in}\cup \rgst{anc}$ satisfies that all the two-qubit gates in $U$ are control-$Z$ gates and each control-$Z$ gate is preceded and followed by two $Z$ gates.
    In addition, the control-$Z$ gates space evenly.
    Let $T_1$ be the set $\{t:U_t\mathrm{\;is\;a\operatorname{one-qubit} gate.}\}$ and $T_2$ be the set $\{t:U_t\mathrm{\;is}\operatorname{control-}Z.\}$
    Let $H$ be a Hamiltonian acting on the Hilbert space $\hs{H}_{clock}=\spn\{\ket{x}\reg{in\cup anc}\otimes\ket{\clock{t}}\reg{clock}:x\in\{0,1\}^{n+n_a},t\in\{0\}\cup [T]\}$, where $\braket{\clock{t}}{\clock{t'}}=\delta_{t,t'}$ for all $t,t'\in\{0\}\cup [T]$. 
    The Hamiltonian $H$ is defined as follows.
    \begin{equation}\label{eq:3local}
        H = (T+1)H_{out} + J_{in}H_{in} + J_{prop2}H_{prop2} + J_{prop1}H_{prop1},
    \end{equation}
    where 
    \begin{align*}
        H_{out}&= \proj{0}\reg{out}\otimes\proj{\clock{T}}\reg{clock},\\
        H_{in}&= \sum_{i=1}^{n_a}\proj{1}\reg{anc[\mathit{i}]}\otimes \proj{\clock{0}}\reg{clock},\\
        H_{prop1} &= \sum_{t\in T_1} H_{prop,t},\\
        H_{prop2} &= \sum_{t\in T_2} H_{qubit,t}+H_{tiem,t},
    \end{align*}
    where
    \begin{equation*}
        H_{prop,t} = \frac{1}{2}(I\otimes\proj{\clock{t}} + I\otimes\proj{\clock{t-1}}-U_t\otimes \ketbra{\clock{t}}{\clock{t-1}} - U_t^\dagger\otimes \ketbra{\clock{t-1}}{\clock{t}}),
    \end{equation*}
    \begin{equation*}
    H_{qubit,t} = 
    \frac{1}{2}(-2\proj{0}_{f_t}-2\proj{0}_{s_t}+\proj{1}_{f_t}+\proj{s_t})
    \otimes
    (\ketbra{\clock{t}}{\clock{t-1}}+\ketbra{\clock{t-1}}{\clock{t}}),
    \end{equation*}
    where $f_t$ and $s_t$ are the first qubit and the second qubit of the control-$Z$ gate at the $t$th time step, and 
\begin{align*}
    H_{time,t} = \frac{1}{8}I\otimes (&\proj{\clock{t}}+6\proj{\clock{t+1}}+\proj{\clock{t+2}}\\
    &+2\ketbra{\clock{t+2}}{\clock{t}}+2\ketbra{\clock{t}}{\clock{t+2}}\\
    &+\ketbra{\clock{t+1}}{\clock{t}}+\ketbra{\clock{t}}{\clock{t+1}}
    +\ketbra{\clock{t+2}}{\clock{t+1}}+\ketbra{\clock{t+1}}{\clock{t+2}}\\
    &+\proj{\clock{t-3}}+6\proj{\clock{t-2}}+\proj{\clock{t-1}}\\
    &+2\ketbra{\clock{t-1}}{\clock{t-3}}+2\ketbra{\clock{t-3}}{\clock{t-1}}\\
    &+\ketbra{\clock{t-2}}{\clock{t-3}}+\ketbra{\clock{t-3}}{\clock{t-2}}
    +\ketbra{\clock{t-1}}{\clock{t-2}}+\ketbra{\clock{t-2}}{\clock{t-1}}).
\end{align*}
There exist $J_{in}, J_{prop1}, J_{prop_2}\in\poly(n)$ such that if $\Pr[1\gets U(\ket{\psi})]\le \varepsilon$ for all $\ket{\psi}\in\mathbb{C}^{2^n}$, then $\lambda(H)\ge \frac{5}{8}-\varepsilon$.
\end{lemma}

Note that for the Hamiltonian $H_{time,t}$ in Lemma  \ref{lem:kkr}, we need to forward and backward the clock states by one step and two steps.
Because the adjacent vertices in the Johnson graph differ by exactly one element, we can use a 2-local operator to update our clock state by one step.
For the two steps updating, we need the following lemma.

\begin{lemma}\label{lem:path_in_Johnson}
    For any $n\in\mathbb{N}$, there is a Hamiltonian path $(S_1,S_2,\dots, S_T)$ where $T=\binom{n}{2}$ in the Johnson graph $J(n,2)$ such that for all $t\in\{0\}\cup [T-2]$, it holds that 
    $|S_t\cap S_{t+2}|=1$.
\end{lemma}
\begin{proof}
    We find the Hamiltonian path in $J(n,2)$ recursively on $n$.
    We divide the vertices in $J(n,2)$ into two subsets.
    Subset 1 consists of all the vertices that do not include $n$, and Subset 2 consists of $\{1,n\}, \{2,n\},\dots,\{n-1,n\}$.
    Subset 1 forms a Johnson graph $J(n-1,2)$, and Subset 2 forms a clique.
    We first find a path in Subset 1, and then append the vertices in Subset 2.
    \begin{itemize}
        \item Base: $n=3$. output a Hamiltonian path $P_3 = (S_0,S_1,S_2)$ where $S_0=\{1,2\}$, $S_1=\{2,3\}$, and $S_2=\{1,3\}$.
        \item Inductive steps:
        \begin{enumerate}
            \item Find a Hamiltonian path $P_{n-1}$ in $J(n-1,2)$.
    Let $P_{n-1}=(S_0,S_1,\dots,S_{T_n-1},S_{T_n})$ and $S_{T_n}-\{x,n-1\}$ where $x,y\in[n-2]$.
    \item Append $\{n-1,n\}$ to the path.
    \item Append $\{x,n\}$ to the path.
    \item Append $\{z,n\}$ to the path lexicographically, where $z\in[n-2]\setminus\{x\}$.
        \end{enumerate}
    \end{itemize}
    Then we prove $|S_t\cap S_{t+2}|$ for all $t\in[T-2]$ by induction.
    Let $T_n:=\binom{n}{2}$ for all $n\in\mathbb{N}$.
    The base case $n=3$ holds.
    In the inductive step $n$, 
    for all $t>T_{n-1}$, $S_{t}$ contains $n$.
    Hence $|S_t\cap S_{t+2}|=1$ for all $t\in\{T_{n-1}+1, T_{n-1}+2,\dots,T_{n-1}+n=T_{n}\}$.
    According to the algorithm, $S_{T_{n}-1}=\{y,n-1\}$, $S_{T_{n}}=\{x,n-1\}$ and $S_{T_{n}+1}=\{n-1, n\}$, $S_{T_{n}+2}=\{x, n\}$. 
    We have $|S_{T_{n}-1}\cap S_{T_{n}+1}|=1$ and $|S_{T_{n}}\cap S_{T_{n}+2}|=1$.
    As a result, for all $t\le T_{n}$, $|S_t\cap S_{t+2}|=1$.
    This finishes the proof.
\end{proof}

Now we are ready to prove Lemma \ref{lem:3local_main}.
\begin{proof}[Proof of Lemma \ref{lem:3local_main}]
    Let $U=V_TV_{T-1}\dots V_1$ satisfying the structure in Lemma \ref{lem:kkr}.
    We have that $T$ is at most five times of $g$.
    Choose $\nclock=o(n)$ such that $\binom{\nclock}{2}>T$.
    Let $\binom{\nclock}{2}=L$
    Let $S_0,S_1,\dots,S_L$ be the Hamiltonian path of $J(\nclock,2)$ that described in Lemma \ref{lem:path_in_Johnson} and $\{\ket{\clock{t}}\}_{t=0}^{L}$ be the corresponding clock state.

    We construct the Hamiltonian $H$ as follows.
    \begin{equation}
        H = (T+1)H_{out} + J_{in}H_{in} + J_{prop1}H_{prop1} + J_{prop2}H_{prop2}+J_{stab}H_{stab},
    \end{equation}
    where 
    \begin{align*}
        H_{out}&= \proj{0}\reg{out}\otimes\proj{11}\reg{S_T},\\
        H_{in}&= \sum_{i=1}^{n_a}\proj{1}\reg{anc[\mathit{i}]}\otimes \proj{11}\reg{S_0},\\
        H_{prop1} &= \sum_{t\in T_1} H_{prop,t},\\
        H_{prop2} &= \sum_{t\in T_2} H_{qubit,t}+H_{tiem,t},\\
        H_{stab} &= H_{>2}+H_{<2}+H_{t>T}-\bigg({\binom{\nclock}{2}-1}\bigg)I,
    \end{align*}
    where
    \begin{align*}
        H_{prop,t} =& \frac{1}{2}(I\otimes\proj{11}\reg{S_t} + I\otimes\proj{11}\reg{S_{t-1}}\\
        & -U_t\otimes \ketbra{0}{1}\reg{S_{t-1}\setminus S_{t}}\otimes \ketbra{1}{0}\reg{S_{t}\setminus S_{t-1}}
        - U_t^\dagger\otimes \ketbra{1}{0}\reg{S_{t-1}\setminus S_{t}}\otimes \ketbra{0}{1}\reg{S_{t}\setminus S_{t-1}},
    \end{align*}
    \begin{align*}
    H_{qubit,t} = 
    &\frac{1}{2}(-2\proj{0}_{f_t}-2\proj{0}_{s_t}+\proj{1}_{f_t}+\proj{s_t})
    \otimes\\
    &(\ketbra{0}{1}\reg{S_{t-1}\setminus S_{t}}\otimes \ketbra{1}{0}\reg{S_{t}\setminus S_{t-1}}+\ketbra{1}{0}\reg{S_{t-1}\setminus S_{t}}\otimes \ketbra{0}{1}\reg{S_{t}\setminus S_{t-1}}),
    \end{align*}
    where $f_t$ and $s_t$ are the first qubit and the second qubit of the control-$Z$ gate at the $t$th time step, and 
\begin{align*}
    H_{time,t} = \frac{1}{8}I\otimes (&\proj{11}\reg{S_t}+6\proj{11}\reg{S_{t+1}}+\proj{11}\reg{t+2}\\
    &+2\ketbra{0}{1}\reg{S_{t}\setminus S_{t+2}}\otimes \ketbra{1}{0}\reg{S_{t+2}\setminus S_{t}}
    +2\ketbra{0}{1}\reg{S_{t+2}\setminus S_{t}}\otimes \ketbra{1}{0}\reg{S_{t}\setminus S_{t+2}}\\
    &+\ketbra{0}{1}\reg{S_{t}\setminus S_{t+1}}\otimes \ketbra{1}{0}\reg{S_{t+1}\setminus S_{t}}
    +\ketbra{0}{1}\reg{S_{t+1}\setminus S_{t}}\otimes \ketbra{1}{0}\reg{S_{t}\setminus S_{t+1}}\\
    &+\ketbra{0}{1}\reg{S_{t+1}\setminus S_{t+2}}\otimes \ketbra{1}{0}\reg{S_{t+2}\setminus S_{t+1}}
    +\ketbra{0}{1}\reg{S_{t+2}\setminus S_{t+1}}\otimes \ketbra{1}{0}\reg{S_{t+1}\setminus S_{t+2}}\\
    &+\proj{11}\reg{S_{t-3}}+6\proj{11}\reg{S_{t-2}}+\proj{11}\reg{t-1}\\
    &+2\ketbra{0}{1}\reg{S_{t-1}\setminus S_{t-3}}\otimes \ketbra{1}{0}\reg{S_{t-3}\setminus S_{t-1}}
    +2\ketbra{0}{1}\reg{S_{t-3}\setminus S_{t-1}}\otimes \ketbra{1}{0}\reg{S_{t-1}\setminus S_{t-3}}\\
    &+\ketbra{0}{1}\reg{S_{t-2}\setminus S_{t-3}}\otimes \ketbra{1}{0}\reg{S_{t-3}\setminus S_{t-2}}
    +\ketbra{0}{1}\reg{S_{t-2}\setminus S_{t-3}}\otimes \ketbra{1}{0}\reg{S_{t-3}\setminus S_{t-2}}\\
    &+\ketbra{0}{1}\reg{S_{t-1}\setminus S_{t-2}}\otimes \ketbra{1}{0}\reg{S_{t-2}\setminus S_{t-1}}
    +\ketbra{0}{1}\reg{S_{t-2}\setminus S_{t-1}}\otimes \ketbra{1}{0}\reg{S_{t-1}\setminus S_{t-2}}),
\end{align*}
and
\begin{align*}
    H_{>2} &:=\binom{\nclock}{2}\sum_{\rgst{S}:|\rgst{S}|=3} \proj{111}\reg{S},\\
    H_{<2} &:=\sum_{\rgst{S}:|\rgst{S}|=2}\sum_{x\in\{0,1\}^{2}\setminus\{11\}} \proj{x}\reg{S},\\
    H_{t>T} &:= \sum_{t>T}\proj{11}\reg{S_{t}}.
\end{align*}
We can see that $H$ is 3-local.

For the yes case, if there is a state $\ket{\psi}$ accepted by $U$ with probability $1-\varepsilon$,
then the state $\ket{hist(\psi)}$ defined in \Cref{eq:hist} satisfies that 
\begin{align*}
    \expec{hist(\psi)}{H_{in}} &= 0,\\
    \expec{hist(\psi)}{H_{prop1}} &= 0,\\
    \expec{hist(\psi)}{H_{prop2}} &=0,\\
    \expec{hist(\psi)}{H_{stab}} &=0,\\
\end{align*}
and
\begin{equation*}
    \expec{hist(\psi)}{H_{out}} =\frac{\varepsilon}{T+1}.
\end{equation*}
This finishes the first part of the proof.

Then,
we are going to apply projection lemma on $H$.
Let $H_1= (T+1)H_{out}+J_{in}H_{in} + J_{prop1}H_{prop1}+J_{prop2}H_{prop2}$, $H_2=J_{stab}H_{stab}$ and
 $\hs{S}=\spn\{\ket{x}\reg{in\cup anc}\otimes\ket{\clock{t}}\reg{clock}:x\in\{0,1\}^{n+n_a},t\in\{0\}\cup [T]\}$.
 We have that $\hs{S}$ is the zero eigenspace of $H_{stab}$.
 When $J_{in}, J_{prop1}, J_{prop2}\in\poly(n)$, we have $\|H_1\|\le\poly(n)$ by triangular inequality.
The smallest non-zero eigen value of $H_{stab}$ is 1.
Hence, we can choose $J_{stab} = \poly(n)$ such that $\frac{\|H_1\|^2}{J-\|H_1\|}<\frac{1}{8}$.
By projection lemma, we have $\lambda(H)\ge \lambda(H_1\vert_{\hs{S}})-\frac{\|H_1\|^2}{J=\|H-1\|}\ge \lambda(H_1\vert_{\hs{S}})-\frac{1}{8}$.
Also, we have that $\lambda(H_1\vert_{\hs{S}})$ equals to $H$ defined in Lemma \ref{lem:kkr}.
As a result, we have $\lambda(H) \ge \lambda(H_1\vert_{\hs{S}})-\frac{1}{8}\ge \frac{5}{8}-\varepsilon-\frac{1}{8}=\frac{1}{2}-\varepsilon$.
This finishes the proof.
\end{proof}

\begin{corollary}
    The local Hamiltonian problem for 3-local Hamiltonians cannot be solved in $O(2^{\frac{n}{2}(1-\varepsilon)})$ time for any quantum algorithm for any $\varepsilon>0$ if QSETH holds.
\end{corollary}

\begin{corollary}
    Approximating the quantum partition function for 3-local Hamiltonians for any constant relative error cannot be solved in $O(2^{\frac{n}{2}(1-\varepsilon)})$ for any quantum algorithm for any $\varepsilon>0$ time if QSETH holds.   
\end{corollary}

%% file: 5_algorithm.tex
\section{A $O^{\ast}(2^{\frac{n}{2}})$-time algorithm for $k$QPF problem}
In this section, we propose an algorithm that solves $kQPF(H, \beta, \delta)$ in $O^{\ast}(2^{\frac{n}{2}})$ time.
The input $H$ a constant-local, semi-positive $n$-qubit Hamiltonian that satisfies $\|H\|< 1$.
Let $E_p$ be the eigenvalue of $H$ and $\ket{\psi_p}$ be the corresponding eigenstate for each $p\in [2^{n}]$.
The inverse temperature $\beta<\poly(n)$.
The error parameter $\delta\ge\frac{1}{2}+\frac{1}{p(n)}$ where $p(\cdot)$ is an arbitrary polynomial.

The idea is to evenly divide the energy range $[0,1)$ into $T=\poly(n)$ intervals $\{I_j\}_{j=1}^{T}$, where $I_j=[\frac{j-1}{T},\frac{j}{T})$. Let $\Delta:=\frac{\beta}{T}$.
Then, we estimate the number of eigenstates $M_j$ inside each interval $I_j$, and let  $\widetilde{M}_j$ be the estimation. 
Then, we estimate the partition function as follows
\begin{equation}\label{eq:esti_Z}
    \widetilde{Z}:= \sum_{j=1}^{T} \widetilde{M}_j e^{-(j-1)\Delta}.
\end{equation}

\begin{lemma}[Approximating partition function by counting]
\label{lem:qacqpf}
    Let $\widetilde{M}_j$ be the estimation of the number of eigenstates such that   
    for each $j\in[T]$,
    \begin{equation}
    \label{eq:51}
    (1-\frac{1}{4^c})M_j \le \widetilde{M}_j \le (1+\frac{1}{4^c})M_{I'_j}, 
\end{equation}
where $I'_j:=\big[(j-1-\frac{1}{2})\frac{1}{T}, (j+\frac{1}{2})\Delta\big)$ and $M_{I'_j}$ is the exact number of eigenstates inside $I'_j$, then,
\begin{equation}
    \label{eq:52}
    \big(1-(\frac{1}{2}+\frac{1}{n^c})\big)Z\le\frac{\widetilde{Z}}{2} \le \big(1+(\frac{1}{2}+\frac{1}{n^c})\big)Z.    
\end{equation}
\end{lemma}
\begin{proof}
    By \cite{BCGW22}, if 
    \begin{equation}
    \label{eq:requirement_of_DOS}
    (1-\delta_{M})M_j \le \widetilde{M}_j \le (1+\delta_{M})M_{I'_j}, 
\end{equation}
 holds, then we have
    \begin{equation}
    (1-\delta_{M})Z\le \widetilde{Z} \le (1+\delta_{M})(1+e^{\Delta}+e^{2\Delta})Z.
\end{equation}
Choosing $\delta_{M}=\frac{1}{4n^c}$ and $\Delta=\frac{1}{4n^c}$, we have $e^{\Delta}=1+\frac{1}{4n^c}+O(\frac{1}{n^{2c}})$. Therefore $1+e^{\Delta}+e^{2\Delta}= 3 + \frac{3}{4n^c}+O(\frac{1}{n^{2c}})$.
We get
\begin{equation}
    \label{eq:estimate_of_Z}
    (1-\frac{1}{4n^c})Z\le\widetilde{Z} \le (3+\frac{7}{4n^c})Z.    
\end{equation}
Divide \Cref{eq:estimate_of_Z} by 2, we have get\begin{equation}
    \label{eq:estimate_of_Z_2}
    \big(1-(\frac{1}{2}+\frac{1}{8n^c})\big)Z\le\frac{\widetilde{Z}}{2} \le \big(1+(\frac{1}{2}+\frac{7}{8n^c})\big)Z,    
\end{equation}
which satisfies \Cref{eq:52}.
\end{proof}

As a result, we output $\frac{\widetilde{Z}}{2}$ as the estimation of the partition function that has relative error within $1\pm\delta$ satisfying $\delta=\frac{1}{2}+\frac{1}{n^c}$.

Before explaining how to implement the estimation of $M_j$, we first present the tools we are going to use: the phase estimation and the quantum counting.

The goal of phase estimation is to find the eigenvalue of a unitary.
Given the description of a unitary $U$ associated with its eigenstate $\ket{u_\theta}$ satisfying $U\ket{u_\theta}=e^{i\theta}\ket{u_\theta}$, a phase estimation algorithm is to output $\widetilde{\theta}$ that estimate $\theta$ up to an additive error with sufficient successful probability.

\begin{lemma}[Phase estimation and its performance\cite{NC10}]
    \label{lem:phase_estimation}
    Let $P_{U,\ell}$ \yc{P for phase estimation.} be the circuit that executes phase estimation for a $n$-qubit unitary $U$. 
    (The parameter $\ell$ will be explained later.)
    The circuit $P_{U,\ell}$ acts on $\rgst{C}\cup \rgst{T}$ where $|\rgst{C}|=\ell$ that determines the running time, precession, and successful probability,  and $|\rgst{T}|=n$.
    Let $CU^r$ denote the control-$U$ operation.
    The construction of $P_{U,\ell}$  is described below.
    \begin{equation}\label{eq:PE_circuit}
        P_{U,\ell}:=QFT^\dagger\reg{C} \cdot CU^{2^{\ell-1}}\reg{C[1]\cup T} \cdot CU^{2^{\ell-2}}\reg{C[2]\cup T} \cdots 
        CU^{2^{0}}\reg{C[\ell]\cup T},
    \end{equation}
    where $QFT^\dagger$ is the inverse quantum Fourier transform operation over $\ell$ qubits. 
    
    Let $\ket{u_\theta}$ be a eigenstate of $U$ such that $U\ket{u_\theta}=e^{i\theta}\ket{u_\theta}$.
    Let $\widetilde{x}\in\{0,1\}^{\ell}$ be the measurement outcome on $\rgst{C}$ of the state $P_{U,\ell}\sum_{x\in\{0,1\}^\ell}\frac{1}{\sqrt{2^\ell}}\ket{x}\reg{C}\ket{u_\theta}\reg{T}$.
    Let $\widetilde{\theta}:=\frac{int(\widetilde{x})}{2^\ell}$.
    
    For any $b<\ell-1$, it holds that
    \begin{equation}\label{eq:performance_of_PE}
        \Pr[|\widetilde{\theta}-\theta|>\frac{2\pi}{2^b}]\le\frac{1}{2^{\ell-b}}.
    \end{equation}
\end{lemma}

When we choose $b=\ell-2$, we have that the outcome $\widetilde{\theta}$ in the interval $\theta\pm\frac{2\pi}{2^b}$ has a probability greater than $\frac{3}{4}$.
We can amplify the probability to $1-\negl(n)$ by the median of means technique.

\begin{lemma}[Confidence amplification of phase amplification]\label{lem:pe_amplification}
    Following Lemma \ref{lem:phase_estimation}, there is a circuit $A_{U,\ell,m}$ \yc{A for amplification} that executes $m$ times of $P_{U,\ell}$ and some postprocessing.
    The output $\widetilde{\theta}$ satisfies that
    \begin{equation}\label{eq:amplification_of_PE}
        \Pr[|\widetilde{\theta}-\theta|>\frac{2\pi}{2^b}]\le 1-e^{-O(m)},
    \end{equation}
    for any $b<\ell-1$
\end{lemma}
\begin{proof}
    We prepare $m$ of many $\ell$-qubit registers $\rgst{C_1}, \rgst{C_2},\dots, \rgst{C_m}$ and set the state in each of them be $\sum_{x\in\{0,1\}^{\ell}}\frac{1}{\sqrt{2^\ell}}\ket{x}$.
    Let $\rgst{T}$ be in $\ket{u_\theta}$ initially.
    Then, we apply $P_{U,\ell}$ sequentially on $\rgst{C_1}\cup \rgst{T}, \rgst{C_2}\cup \rgst{T},\dots, \rgst{C_m}\cup \rgst{T}$. 
    (Note that the state $\ket{u_\theta}$ stays unchanged after applied by $P_{U,\ell}$.)
    
    Let the outcome in each register be $\widetilde{x}_1, \widetilde{x}_2,\dots, \widetilde{x}_m$.
    By Chernoff bound, the event that there are more than half of outcomes that are in the the interval $(\theta\pm\frac{2\pi}{2^b})\cdot 2^\ell$ is at least $1-e^{-O(m)}$.
    Hence, the median of $\widetilde{x}_1, \widetilde{x}_2,\dots, \widetilde{x}_m$ lying in the interval has a probability greater than $1-e^{-O(m)}$.
    We construct $A_{U, \ell, m}$ by appending a circuit that computes the median of them to the end of the circuit that runs the $m$ of  many $P_{U,\ell}$.
\end{proof}

Next, we present the second tool: the quantum counting, which is also known as the amplitude estimation.
Let the final state of a quantum circuit is in the superposition of a ``good state'' and a ``bad state''.
The goal of the quantum counting, or, the amplitude estimation, is to estimate the amplitude of the ``good state.  
We can use phase estimation to implement the quantum counting.

\begin{lemma}[Quantum counting \cite{NC10}]
    Let a quantum circuit $U$ taking a $(n+1)$-qubit input state $\ket{\psi}$ satisfy $U\ket{\psi}=\sqrt{\frac{M}{2N}}\ket{0}\reg{out}\ket{\xi_0}\reg{\overline{out}} + \sqrt{\frac{N-M}{2N}}\ket{1}\reg{out}\ket{\xi_1}\reg{\overline{out}}$,
    where $N=2^n$ and $M\le N$.
    If $U$ runs in $\poly(n)$ time and $\ket{\psi}$ can be prepared in $\poly(n)$ time,
    then there is a quantum circuit that runs in $O(\poly(n)\cdot 2^{\frac{n}{2}})$ times outputs $\widetilde{M}$ such that $(1-\frac{1}{n^c})M\le\widetilde{M}\le(1+\frac{1}{n^c})M$ with probability $1-\negl(n)$.
\end{lemma}\
There has already been a proof of $(1-\frac{1}{\sqrt{M}})M\le\widetilde{M}\le (1+\frac{1}{\sqrt{M}})M$ in \cite{NC10}.
However, we need the relative error to be $\frac{1}{n^c}$.
The proof is almost the same in \cite{NC10}.
For completeness, we write down proof here.
\begin{proof}
    Let $G:=U(2\proj{\psi}-I)U^\dagger(2\proj{0}\reg{out}-I)$.
    It holds that when $G$ is restricted on the two-dimension subspace $\spn(\{\ket{1}\ket{\xi_1}, \ket{0}\ket{\xi_0}\})$, the eigenvalues are $e^{i\theta}$ and $e^{i(2\pi-\theta)}$
    where $\theta\in[0,2\pi)$ satisfies $\sin \frac{\theta}{2}=\sqrt{\frac{M}{N}}$. 
    Therefore, we can execute phase estimation for $G$ on the input state $U\ket{\psi}$ to find $\theta$.

    We apply $A_{G,\ell,m}$ (defined in Lemma \ref{lem:pe_amplification}) on the input state $U\ket{\psi}$. 
    Let the outcome be $\theta'$ and $
    \widetilde{\theta}:=\min\{\theta',2\pi-\theta'\}$,
    and let $\Delta \theta:=|\widetilde{\theta}-\theta|$, $\widetilde{M}:=N\sin^2\frac{\widetilde{\theta}}{2}$.
    It holds that 
    \begin{equation}\label{eq:quantum_counting_error}
        |\widetilde{M}-M|<(\sqrt{2NM}+\frac{N\Delta\theta}{2})\Delta\theta.
    \end{equation}
    When we choose $\ell=\frac{n}{2}+c\log n +2$, where $c$ is a constant.
    Let $b=\frac{n}{2}+c\log n$ (where $b$ is the parameter in \Cref{eq:amplification_of_PE}). 
    We have $\Delta\theta = \frac{1}{2^b} = \frac{1}{2^\frac{n}{2}\cdot n^c}$.
    By \Cref{eq:quantum_counting_error} and \Cref{eq:amplification_of_PE}, we have
    $|\widetilde{M}-M|< O(\frac{1}{n^c}\sqrt{M})$ with probability $1-e^{-O(m)}$.
    As a result, the probability that $(1-\frac{1}{n^c})M\le\widetilde{M}\le(1+\frac{1}{n^c})M$ is $1-\negl(n)$ when we choose $m=\poly(n)$.

    The execution that dominates the running time in $A_{G,\ell,m}$ is $CG^{2^\ell}$ where $CG^{2^\ell}$ is control-$G^{2^\ell}$ and $\ell=\frac{n}{2}c\log n +2$, and there are $m=\poly(n)$ many of $CG^{2^\ell}$.
    We assume that the subroutines inside $G$, that is, $U$, $U^\dagger$, and $2\proj{\psi}-I$ are $\poly$ time. 
    Hence, the running time of $A_{G,\ell,m}$ is $O(\poly(n)\cdot 2^{\frac{n}{2}})$.
\end{proof}

Now we explain the idea of how to estimate $M_j$, the number of eigenstates in the interval $I_j=\big[\frac{(j-1)}{T}, \frac{j}{T}\big)$, for each $j$.
We try to construct a circuit $U_j$ that verifies whether an eigenvalue $E_p$ is in the interval $I_j$.
We apply $U_j$ on the uniform superposition of all the eigenstates.
The final state is the superposition of the ``good state'', which corresponds to the energies in $I_j$, and the ``bad state'', which corresponds to the energies not in $I_j$.
By running quantum counting, we can estimate $M_j$.

An issue we encounter is that the eigenstates of $H$ are unknown.
To overcome this issue, we use the fact that the uniform superposition over a complete basis tensor product with its complex conjugate is identical to an EPR state. 
Therefore, we can apply $U_j$ on the first half of EPR state and then execute the quantum counting. 

\begin{lemma}\label{lem:choi}
    For any complete orthogonal basis of $n$-qubit system $\{\ket{\psi_i}\}_{i=1}^{N}$,i.e., $\sum_{p=1}^{N}\proj{\psi_p}=I$ and $\inner{\psi_p}{\psi_{p'}}=\delta_{p,p'}$, it holds that
    \begin{equation}
        \sum_{p=1}^{N}\ket{\psi_p}\otimes\ket{\psi_p^{\ast}} = \sum_{q=1}^{N}\ket{bin(q)}\otimes\ket{bin(q)},
    \end{equation}
    where $\ket{\psi}$ is the complex conjugate of $\ket{\psi_p}$.
\end{lemma}
\begin{proof}
    Let $V=\sum_{p,q=1}^{N}V_{q,p}\ketbra{q}{p}$ be a unitary such that $\ket{\psi_p}= V\ket{p}$ for all $p\in[N]$.
    Here we use $\ket{p}$ as a shorthand notation for $\ket{bin{(p)}}$.
    We have 
    \begin{align}
        \sum_{p=1}^{N}\ket{\psi_p}\otimes\ket{\psi_p^{\ast}} 
        & = \sum_{p=1}^{N}\bigg(\Big(\sum_{q=1}^{N}V_{q,p}\ket{q}\Big)\otimes\Big(\sum_{q'=1}^{N}V_{q'p}^{\ast}\ket{q'}\Big)\bigg)\nonumber\\
        & =  \sum_{p=1}^{N}\sum_{q=1}^{N}\sum_{q'=1}^{N}V_{q,p}V^{\ast}_{q'p}\ket{q}\otimes\ket{q'}.\label{eq:Choi_1}
    \end{align}
    Because $V$ is a unitary, we have $\sum_{p=1}^{N}V_{qp}V^{\ast}_{q'p}=\delta_{q,q'}$.
    \Cref{eq:Choi_1} becomes $\sum_{q=1}^{N}\ket{q}\otimes\ket{q}$.
\end{proof}

Now we are ready to show our algorithm for approximating the quantum partition function.
\begin{theorem}[$O^{\ast}(2^{\frac{n}{2}})$ time algorithm for $kQPF$]
    \label{thm:main_qpf_alg}
    There exists a quantum algorithm that solves $kQPF(H, \beta, \delta)$ in $O^{\ast}(2^{\frac{n}{2}})$ time, where $H$ is a constant local and semi-definite Hamiltonian acting on $n$ qubits, $\beta<\poly(n)$, and $\delta>\frac{1}{2}+\frac{1}{n^c}$ for arbitrary constant $c$ with successful probability $1-\negl(n)$.  
\end{theorem}
\begin{proof}
    We write down the algorithm.
    \begin{enumerate}
        \item Let $T=4n^c$ and $\Delta=\frac{\beta}{T}$. For $j\in[T]$:
        \begin{enumerate}
            \item Prepare the EPR state $\ket{\Psi}:=\sum_{q\in[2N]}\ket{bin(q)}\ket{bin(q)}$.
            \item Execute quantum counting on $U_j\ket{\Psi}$, where $U_j$ acts on the first half of the EPR state.
            The construction of $U_j$ will be explained later.
            Let $\widetilde{M}_j$ be the output of the quantum counting.
        \end{enumerate}
        \item Let $\zout:=\sum_{j\in[T]}\frac{1}{2}\widetilde{M}_je^{-(j-1)\Delta}$.
        \item Output $\zout$.
    \end{enumerate}

    The circuit $U_j:=U_{dec}U_{EE}$ that verifies whether $E_p$ in $I_j$.
    
    The energy estimation circuit $U_{EE}$ takes an eigenstate $\ket{\psi_p}$ and outputs the corresponding energy $E_p$ up to an additive error $\frac{\varepsilon}{2}$ with successful probability $1-\eta$.
    The circuit $U_{dec}$ decides if the output of $U_{EE}$ in the interval $I_j$.

    The energy estimation $U_{EE}$ satisfies that for all $p\in[N]$ (where $N=2^n$),
\begin{equation}\label{eq:EE}
    U_{EE}\ket{\psi_p} = \sum_{E'}\alpha_{E'}\ket{bin(E')}\ket{\phi_p},
\end{equation}
where $\sum_{E':E'\in [E-\frac{\varepsilon}{2}, E+\frac{\varepsilon}{2})}\alpha_{E'}\ge 1-\eta$, and $\ket{\phi_p}$ is some state depending on $\ket{\psi_p}$. 

We add one qubit to extend the range of $p$ to $2N$. 
We have that $U_j$ is $\eta$-close to $U_{j,ideal}$ defined as follows.

\begin{equation}
    \label{eq:Uj}
    U_{j,ideal}\ket{\psi_p}=\left\{
    \begin{array}{ll}
         \ket{1}\ket{\phi_p} &\mathrm{,\;if\;} E_p\in[(j-1)\Delta, j\Delta), \\
         (\alpha_p\ket{1}+\beta_p\ket{0})\ket{\phi_p}& 
         \mathrm{,\;if\;} E_p\in[(j-1)\Delta-\varepsilon, (j-1)\Delta-\frac{\varepsilon}{2})\mathrm{\;or\;}p\in[(j)\Delta+\frac{\varepsilon}{2}, j\Delta+\varepsilon),\\
         \ket{0}\ket{\phi_p} 
         &\mathrm{,\;if\;} E_p\notin[(j-1)\Delta-\varepsilon, j\Delta+\varepsilon)\mathrm{\;or\;}p>N,
    \end{array}
    \right.
\end{equation}
where $\alpha_p,\beta_p$ are some complex numbers satisfying  $|\alpha_p|+|\beta_p|=1$.

When $U_{j,ideal}$ acts on the first half of $\ket{\Psi}$,
combining \Cref{eq:Uj} and Lemma \ref{lem:choi}, 
we have
\begin{equation}
    U_{j,ideal}\ket{\Psi} = \sqrt{\frac{M'_j}{2N}}\ket{1}\ket{\xi'_1} + \sqrt{\frac{2N-M'_j}{2N}}\ket{0}\ket{\xi'_0}, 
\end{equation}
where $M_j \le M'_j \le M^{\varepsilon}_j$, and $M^{\varepsilon}_j$ is defined by the number of eigestates in the interval $[\frac{j-1}{T}-\varepsilon, \frac{j}{T} +\varepsilon)$,
and $\ket{\xi_1}$, $\ket{\xi_2}$ are two quantum states that are orthogonal to each other.

Let $\widetilde{M}_{j,ideal}$ be the output of quantum counting for $U_{j,ideal}\ket{\Psi}$.
It holds that $(1-\frac{1}{n^c})M'\le \widetilde{M}_{j,ideal} \le (1+\frac{1}{n^c})M'_j$ with probability $1-\negl(n)$.
Because $M_j \le M'_j \le M^{\varepsilon}_j$ ,we have
$(1-\frac{1}{n^c})M\le \widetilde{M}_j \le (1+\frac{1}{n^c})M^{\varepsilon}_j$ with probability $1-\negl(n)$.

By replacing $U_{j,ideal}$ with $U_j$ in the circuit of quantum counting one by one, we obtain the quantum counting for $U_{j,ideal}\ket{\Psi}$.
There are $\poly(n)\cdot 2^{\frac{n}{2}}$ if many $U_{j_ideal}$ in the quantum counting. 
By union bound, if $\eta=e^{-O(n^2)}$, then we have that the successful probability of the quantum counting for the real implementation is also $1-\negl(n)$.

As the result, the output $\widetilde{M}_j$ in Step 1.2 satisfies the condition in Lemma \ref{lem:qacqpf} with probability $1-\negl(n)$.
By union bound, $\widetilde{M}_j$ satisfies the condition for all $j\in[T]$ is $1-\negl(n)$ as well.
Consequently, our algorithm approximates $Z$ with a relative error $\frac{1}{2}+\frac{1}{n^c}$ with successful probability $1-negl(n)$.

Finally, we explain how to implement the energy estimation circuit $U_{EE}$.
The ideal is applying the phase estimation to the Hamiltonian evolution for one unit time $e^{-iH}$.
It holds that $e^{-iH}\ket{\psi_p} = e^{-iE_p}\ket{\psi_p}$.
We apply phase estimation for $e^{-iH}$ up to an additive error $\frac{\varepsilon}{2}$ with successful probability $1-\eta$, 
that is, to execute $A_{e^{-iH}, \ell, m}$ on the input $\ket{\psi_p}$.
To achieve the additive error $\frac{\varepsilon}{2}$ and successful probability $1-\eta=1-e^{-O(n^2)}$, we choose $\ell=\log \frac{1}{\varepsilon}+3$ and $m=n^2$.

In the phsae estimation circuit, we need to execute $e^{-iH}, e^{-2iH},\dots, e^{2^{\ell-1}H}$. 
We use a Hamiltonian simulation algorithm for $k$-local Hamiltonian that implements a unitary that is $\xi$-close to $e^{-iHt}$ for all $t$ in $O(\poly(n,t,\log\frac{1}{\xi}))$ time e.g., \cite{LC17}.
Consider the additive error $\frac{\varepsilon}{2}=1/\poly(n)$.
Choose $\xi=e^{-n}. 
$By union bound, the implementation is $\negl(n)$-close to $U_{EE}$. 
The operation dominates the running time of phase estimation is $e^{-2^{\ell-1}iH}$, which can be simulated in $O(\poly(n,2^{\ell-1},\log\frac{1}{\xi})=O(n, O(\frac{1}{\varepsilon}),n)=\poly(n)$ time.
\end{proof}

%% file: A_SAT2H.tex
\section{A trivial fine-grained reduction to $k(\varepsilon)$LH from $k(\varepsilon)$SAT}
\label{sec:appexdixA}
\begin{theorem}[Lower bound of $k$LH]
    \label{thm:almost_make_me_cry_thm}
    Assuming QSETH, for any $\varepsilon > 0$, there is $k$ (depending on $\varepsilon$) such that for any algorithm, there exists $n_0\in\mathbb{N}$ such that for all $n\ge n_0$, there is a Hamiltonian $H$ acting on $n$ qubits, associated $a,b$ satisfying $b-a\ge1/\poly(n)$ such that $LH(H, a, b)$ cannot be solved in $O(2^{\frac{n}{2}(1-\varepsilon)})$ time.
\end{theorem}
\begin{proof}
    We construct a Hamiltonian $H$ from a $k$SAT instance $\Phi=\varphi_1\wedge \varphi_2\wedge\cdots\wedge\varphi_m$.

    Let $S_i\subseteq[n]$ be the set collecting the index $j$ such that $x_j$ or $\neg x_j$ appears in $\varphi_i$, and let $\rgst{S_i}$ be the corresponding register. 
    Let $y_i\in\{0,1\}^{|S_i|}$ be the assignment to the variables appearing in $\varphi_i$ such that $\varphi_i(y_i)=0$.
    Because $\varphi_i$ is in disjunctive form, $y_i$ is unique.
    Let $H:=\sum_{i=1}^{m}H_i$, where $H_i:= I-\proj{y_i}\reg{S_i}$ for all $i\in[m]$.
    It holds that $H_i\ket{y_i}=0$ if $\varphi_i(y_i)=1$ and $H_i\ket{y_i}=\ket{y_i}$ if $\varphi_i(y_i)=0$.
    Each $H_i$ acts non-trivially on at most $k$ qubits.
    Hence, $H$ is $k$-local.
    
    If there exist an assignment $x\in\{0,1\}^n$ satisfying $\Phi$, then $H\ket{x}=0$. Otherwise, $\lambda(H)\ge 1$.
    Hence, solving $LH(H, a=1/n,b=1-1/n)$ can decide $k$SAT.
    The construction of $H$ takes $\poly(n)$ time.
    Hence, if $LH(H, a=1/n,b=1-1/n)$ is solved in $O(2^{\frac{n}{2}}(1-\varepsilon))$ time, then $kSAT(\Phi)$ is also solved in $O(2^{\frac{n}{2}}(1-\varepsilon))$ time, which violates QSETH.
\end{proof}

%% file: main.bbl
\newcommand{\etalchar}[1]{$^{#1}$}
\begin{thebibliography}{KGM{\etalchar{+}}24}

\bibitem[ACL{\etalchar{+}}20]{ACL+20}
Scott Aaronson, Nai-Hui Chia, Han-Hsuan Lin, Chunhao Wang, and Ruizhe Zhang.
\newblock On the quantum complexity of closest pair and related problems.
\newblock In {\em Proceedings of the 35th Computational Complexity Conference},
  CCC '20, Dagstuhl, DEU, 2020. Schloss Dagstuhl--Leibniz-Zentrum fuer
  Informatik.
\newblock \href {https://arxiv.org/abs/1911.01973} {\path{arXiv:1911.01973}},
  \href {https://doi.org/10.4230/LIPIcs.CCC.2020.16}
  {\path{doi:10.4230/LIPIcs.CCC.2020.16}}.

\bibitem[AGIK09]{AGIK09}
Dorit Aharonov, Daniel Gottesman, Sandy Irani, and Julia Kempe.
\newblock The power of quantum systems on a line.
\newblock {\em Communications in Mathematical Physics}, 287(1):41–65, January
  2009.
\newblock \href {https://arxiv.org/abs/0705.4077} {\path{arXiv:0705.4077}},
  \href {https://doi.org/10.1007/s00220-008-0710-3}
  {\path{doi:10.1007/s00220-008-0710-3}}.

\bibitem[Als12]{Als12}
Brian Alspach.
\newblock Johnson graphs are {H}amilton-connected.
\newblock {\em Ars Mathematica Contemporanea}, 6(1):21--23, 2012.
\newblock \href {https://doi.org/10.26493/1855-3974.291.574}
  {\path{doi:10.26493/1855-3974.291.574}}.

\bibitem[BCGW22]{BCGW22}
Sergey Bravyi, Anirban Chowdhury, David Gosset, and Pawel Wocjan.
\newblock Quantum {H}amiltonian complexity in thermal equilibrium.
\newblock {\em Nature Physics}, 18(11):1367–1370, October 2022.
\newblock \href {https://arxiv.org/abs/2110.15466} {\path{arXiv:2110.15466}},
  \href {https://doi.org/10.1038/s41567-022-01742-5}
  {\path{doi:10.1038/s41567-022-01742-5}}.

\bibitem[BGL{\etalchar{+}}25]{BGL+25}
Harry Buhrman, Sevag Gharibian, Zeph Landau, François Le~Gall, Norbert Schuch,
  and Suguru Tamaki.
\newblock {Beating the Natural Grover Bound for Low-Energy Estimation and State
  Preparation}.
\newblock {\em Physical Review Letters}, 135(3), July 2025.
\newblock \href {https://arxiv.org/abs/2407.03073} {\path{arXiv:2407.03073}},
  \href {https://doi.org/10.1103/29qw-bssx} {\path{doi:10.1103/29qw-bssx}}.

\bibitem[BPS21]{BPS21}
Harry Buhrman, Subhasree Patro, and Florian Speelman.
\newblock {A Framework of Quantum Strong Exponential-Time Hypotheses}.
\newblock In Markus Bl\"{a}ser and Benjamin Monmege, editors, {\em 38th
  International Symposium on Theoretical Aspects of Computer Science (STACS
  2021)}, volume 187 of {\em Leibniz International Proceedings in Informatics
  (LIPIcs)}, pages 19:1--19:19, Dagstuhl, Germany, 2021. Schloss Dagstuhl --
  Leibniz-Zentrum f{\"u}r Informatik.
\newblock \href {https://arxiv.org/abs/1911.05686} {\path{arXiv:1911.05686}},
  \href {https://doi.org/10.4230/LIPIcs.STACS.2021.19}
  {\path{doi:10.4230/LIPIcs.STACS.2021.19}}.

\bibitem[CCH{\etalchar{+}}23]{CCH+23}
Nai-Hui Chia, Kai-Min Chung, Yao-Ching Hsieh, Han-Hsuan Lin, Yao-Ting Lin, and
  Yu-Ching Shen.
\newblock {On the Impossibility of General Parallel Fast-Forwarding of
  Hamiltonian Simulation}.
\newblock In {\em Proceedings of the 38th Computational Complexity Conference},
  CCC '23, Dagstuhl, DEU, 2023. Schloss Dagstuhl--Leibniz-Zentrum fuer
  Informatik.
\newblock \href {https://arxiv.org/abs/2305.12444} {\path{arXiv:2305.12444}},
  \href {https://doi.org/10.4230/LIPIcs.CCC.2023.33}
  {\path{doi:10.4230/LIPIcs.CCC.2023.33}}.

\bibitem[CCK{\etalchar{+}}25]{CCK+23}
Yanlin Chen, Yilei Chen, Rajendra Kumar, Subhasree Patro, and Florian Speelman.
\newblock {QSETH} strikes again: Finer quantum lower bounds for lattice
  problem, strong simulation, hitting set problem, and more.
\newblock In Alina Ene and Eshan Chattopadhyay, editors, {\em Approximation,
  Randomization, and Combinatorial Optimization. Algorithms and Techniques,
  {APPROX/RANDOM} 2025, August 11-13, 2025, Berkeley, CA, {USA}}, volume 353 of
  {\em LIPIcs}, pages 6:1--6:24. Schloss Dagstuhl - Leibniz-Zentrum f{\"{u}}r
  Informatik, 2025.
\newblock \href {https://arxiv.org/abs/arXiv:2309.16431}
  {\path{arXiv:arXiv:2309.16431}}, \href
  {https://doi.org/10.4230/LIPICS.APPROX/RANDOM.2025.6}
  {\path{doi:10.4230/LIPICS.APPROX/RANDOM.2025.6}}.

\bibitem[Cha24]{KLC24}
Garnet Kin-Lic Chan.
\newblock Spiers memorial lecture: Quantum chemistry{,} classical heuristics{,}
  and quantum advantage.
\newblock {\em Faraday Discuss.}, 254:11--52, 2024.
\newblock \href {https://arxiv.org/abs/2407.11235} {\path{arXiv:2407.11235}},
  \href {https://doi.org/10.1039/D4FD00141A} {\path{doi:10.1039/D4FD00141A}}.

\bibitem[GHLS15]{GHLS15}
Sevag Gharibian, Yichen Huang, Zeph Landau, and Seung~Woo Shin.
\newblock Quantum {Hamiltonian} {Complexity}.
\newblock {\em Foundations and Trends® in Theoretical Computer Science},
  10(3):159--282, 2015.
\newblock arXiv: 1401.3916.
\newblock \href {https://arxiv.org/abs/1401.3916} {\path{arXiv:1401.3916}},
  \href {https://doi.org/10.1561/0400000066} {\path{doi:10.1561/0400000066}}.

\bibitem[GP19]{GP19}
Sevag Gharibian and Ojas Parekh.
\newblock Almost optimal classical approximation algorithms for a quantum
  generalization of max-cut.
\newblock In Dimitris Achlioptas and L{\'{a}}szl{\'{o}}~A. V{\'{e}}gh, editors,
  {\em Approximation, Randomization, and Combinatorial Optimization. Algorithms
  and Techniques, {APPROX/RANDOM} 2019, September 20-22, 2019, Massachusetts
  Institute of Technology, Cambridge, MA, {USA}}, volume 145 of {\em LIPIcs},
  pages 31:1--31:17. Schloss Dagstuhl - Leibniz-Zentrum f{\"{u}}r Informatik,
  2019.
\newblock \href {https://arxiv.org/abs/1909.08846} {\path{arXiv:1909.08846}},
  \href {https://doi.org/10.4230/LIPICS.APPROX-RANDOM.2019.31}
  {\path{doi:10.4230/LIPICS.APPROX-RANDOM.2019.31}}.

\bibitem[Gro96]{Gro96}
Lov~K. Grover.
\newblock A fast quantum mechanical algorithm for database search.
\newblock In {\em Proceedings of the Twenty-Eighth Annual ACM Symposium on
  Theory of Computing}, STOC '96, page 212–219, New York, NY, USA, 1996.
  Association for Computing Machinery.
\newblock \href {https://arxiv.org/abs/quant-ph/9605043}
  {\path{arXiv:quant-ph/9605043}}, \href
  {https://doi.org/10.1145/237814.237866} {\path{doi:10.1145/237814.237866}}.

\bibitem[HKW24]{HKW24}
Jeremy~Ahrens Huang, Young~Kun Ko, and Chunhao Wang.
\newblock On the (classical and quantum) fine-grained complexity of
  log-approximate {CVP} and {Max-Cut}, 2024.
\newblock \href {https://arxiv.org/abs/2411.04124} {\path{arXiv:2411.04124}}.

\bibitem[HNN13]{HNN13}
Sean Hallgren, Daniel Nagaj, and Sandeep Narayanaswami.
\newblock The local hamiltonian problem on a line with eight states is
  {QMA}-complete.
\newblock {\em Quantum Info. Comput.}, 13(9–10):721–750, September 2013.
\newblock \href {https://arxiv.org/abs/1312.1469} {\path{arXiv:1312.1469}},
  \href {https://doi.org/10.26421/QIC13.9-10-1}
  {\path{doi:10.26421/QIC13.9-10-1}}.

\bibitem[IPZ01]{IPZ99}
Russell Impagliazzo, Ramamohan Paturi, and Francis Zane.
\newblock Which problems have strongly exponential complexity?
\newblock {\em Journal of Computer and System Sciences}, 63(4):512--530, 2001.
\newblock \href {https://doi.org/10.1006/jcss.2001.1774}
  {\path{doi:10.1006/jcss.2001.1774}}.

\bibitem[KGM{\etalchar{+}}24]{KGM+24}
Alex Kerzner, Vlad Gheorghiu, Michele Mosca, Thomas Guilbaud, Federico
  Carminati, Fabio Fracas, and Luca Dellantonio.
\newblock A square-root speedup for finding the smallest eigenvalue.
\newblock {\em Quantum Science and Technology}, 9(4):045025, August 2024.
\newblock \href {https://arxiv.org/abs/2311.04379} {\path{arXiv:2311.04379}},
  \href {https://doi.org/10.1088/2058-9565/ad6a36}
  {\path{doi:10.1088/2058-9565/ad6a36}}.

\bibitem[KKR04]{KKR04}
Julia Kempe, Alexei~Y. Kitaev, and Oded Regev.
\newblock The complexity of the local hamiltonian problem.
\newblock In Kamal Lodaya and Meena Mahajan, editors, {\em {FSTTCS} 2004:
  Foundations of Software Technology and Theoretical Computer Science, 24th
  International Conference, Chennai, India, December 16-18, 2004, Proceedings},
  volume 3328 of {\em Lecture Notes in Computer Science}, pages 372--383.
  Springer, 2004.
\newblock \href {https://arxiv.org/abs/quant-ph/0406180}
  {\path{arXiv:quant-ph/0406180}}, \href
  {https://doi.org/10.1007/978-3-540-30538-5\_31}
  {\path{doi:10.1007/978-3-540-30538-5\_31}}.

\bibitem[KSV02]{KSV02}
A.~Kitaev, A.~Shen, and M.~Vyalyi.
\newblock {\em Classical and {{Quantum Computation}}}, volume~47 of {\em
  Graduate {{Studies}} in {{Mathematics}}}.
\newblock {American Mathematical Society}, {Providence, Rhode Island}, May
  2002.
\newblock \href {https://doi.org/10.1090/gsm/047} {\path{doi:10.1090/gsm/047}}.

\bibitem[LC17]{LC17}
Guang~Hao Low and Isaac~L. Chuang.
\newblock Optimal {H}amiltonian simulation by quantum signal processing.
\newblock {\em Phys. Rev. Lett.}, 118:010501, January 2017.
\newblock \href {https://arxiv.org/abs/1606.02685} {\path{arXiv:1606.02685}},
  \href {https://doi.org/10.1103/PhysRevLett.118.010501}
  {\path{doi:10.1103/PhysRevLett.118.010501}}.

\bibitem[NC10]{NC10}
Michael~A. Nielsen and Isaac~L. Chuang.
\newblock {\em Quantum Computation and Quantum Information: 10th Anniversary
  Edition}.
\newblock Cambridge University Press, 2010.
\newblock \href {https://doi.org/10.1017/CBO9780511976667}
  {\path{doi:10.1017/CBO9780511976667}}.

\bibitem[OT08]{OT08}
Roberto Oliveira and Barbara~M. Terhal.
\newblock The complexity of quantum spin systems on a two-dimensional square
  lattice.
\newblock {\em Quantum Info. Comput.}, 8(10):900–924, November 2008.
\newblock \href {https://arxiv.org/abs/quant-ph/0504050}
  {\path{arXiv:quant-ph/0504050}}, \href {https://doi.org/10.26421/QIC8.10-2}
  {\path{doi:10.26421/QIC8.10-2}}.

\bibitem[Sac11]{SAC11}
Subir Sachdev.
\newblock {\em Quantum {Phase} {Transitions}}.
\newblock Cambridge University Press, Cambridge, UNITED KINGDOM, 2011.
\newblock \href {https://doi.org/10.1017/CBO9780511973765}
  {\path{doi:10.1017/CBO9780511973765}}.

\bibitem[{Vas}15]{VW15}
Virginia {Vassilevska Williams}.
\newblock {Hardness of Easy Problems: Basing Hardness on Popular Conjectures
  such as the Strong Exponential Time Hypothesis (Invited Talk)}.
\newblock In Thore Husfeldt and Iyad~A. Kanj, editors, {\em 10th International
  Symposium on Parameterized and Exact Computation, {IPEC} 2015, September
  16-18, 2015, Patras, Greece}, volume~43 of {\em LIPIcs}, pages 17--29.
  Schloss Dagstuhl - Leibniz-Zentrum f{\"{u}}r Informatik, 2015.
\newblock \href {https://doi.org/10.4230/LIPICS.IPEC.2015.17}
  {\path{doi:10.4230/LIPICS.IPEC.2015.17}}.

\end{thebibliography}
